\documentclass[12pt]{iopart}
\usepackage{graphicx}
\usepackage{subfigure}
\usepackage{epsfig}
\usepackage{epstopdf}
\usepackage{cancel}
\usepackage[dvipsnames]{xcolor}

\begin{document}

\title[Optimality of optical pumping for a closed $\Lambda$-system]{On the optimality of optical pumping for a closed $\Lambda$-system with large decay rates of the intermediate excited state}

\author{Dionisis Stefanatos$^*$ and Emmanuel Paspalakis}

\address{Department of Materials Science, University of Patras, Patras 265 04, Greece}
\ead{$^*$dionisis@post.harvard.edu}
\vspace{10pt}
\begin{indented}
\item[]December 2021
\end{indented}

\begin{abstract}
We use optimal control theory to show that for a closed $\Lambda$-system where the excited intermediate level decays to the lower levels with a common large rate, the optimal scheme for population transfer between the lower levels is actually optical pumping. In order to obtain this result we exploit the large decay rate to eliminate adiabatically the weakly coupled excited state, then perform a transformation to the basis comprised of the dark and bright states, and finally apply optimal control to this transformed system. Subsequently, we confirm the optimality of the optical pumping scheme for the original closed $\Lambda$-system using numerical optimal control. We also demonstrate numerically that optical pumping remains optimal when the decay rate to the target state is larger than that to the initial state or the two rates are not very different from each other. The present work is expected to find application in various tasks of quantum information processing, where such systems are encountered.
\end{abstract}

\section{Introduction}

\label{sec:intro}

The efficient control of quantum systems comprises one of the basic pillars on which relies the successful implementation of modern quantum technologies \cite{Acin18,Alexeev21}. Among the various techniques which have been developed over time, quantum optimal control \cite{Glaser15,Sola18,Stefanatos20,Boscain21} possesses a central role and, along with the method of shortcuts to adiabaticity \cite{David19}, have dominated the latest years. Optimal control has been successfully applied to a plethora of quantum systems, see for example Refs. \cite{Khaneja01,Alessandro01,Boscain02,Sklarz04,Wu06,Caneva09,Salamon09,Lapert10,Wang10,Hegerfeldt13,Jacobs16,Machnes18,Stefanatos20b,Martikyan20,Lokutsievskiy21} just to cite a few.

A quantum system which has attracted considerable attention among the quantum control community is the three-level $\Lambda$-system, where the two lower lying levels $|1\rangle$ and $|3\rangle$ are connected through an excited intermediate level $|2\rangle$, using the pump and Stokes laser fields, respectively. The celebrated stimulated raman adiabatic passage (STIRAP) method \cite{Bergmann98,Kobrak98,Vitanov01,Thanopulos07,Vitanov17,Bergmann19} was developed for the efficient transfer of population from state $|1\rangle$ to state $|3\rangle$ using a counterintuitive pulse-sequence, where the Stokes pulse, connecting levels $|2\rangle$ and $|3\rangle$, is applied before the pump pulse, connecting levels $|1\rangle$ and $|2\rangle$. If the applied fields are slowly varied, the population transfer takes place along the dark eigenstate of the system, which initially coincides with state $|1\rangle$ and finally with state $|3\rangle$. When optimal control theory is applied to the $\Lambda$-system, then the intuitive pulse sequence is obtained \cite{Boscain02}, with the Stokes pulse preceding the pump pulse. The counterintuitive STIRAP pulse-sequence is recovered when the occupation of the intermediate state $|2\rangle$ is penalized in the cost function \cite{Sola99,Kis02,Kumar11,Assemat12,Rat12}, although it was initially believed that it could not be obtained as a solution to an optimal control problem \cite{Band94}.

In the present work we consider a closed $\Lambda$-system, where the excited state $|2\rangle$ decays to states $|1\rangle$ and $|3\rangle$ with equal rates \cite{Vitanov01,Ivanov05}. For this system we show using optimal control theory that, for large values of the common decay rate, the optimal scheme for population transfer from state $|1\rangle$ to $|3\rangle$ is actually optical pumping, where all the available control energy is put to the pump field while the Stokes field is kept to zero. In order to prove this, we exploit the large decay rate to eliminate adiabatically the weakly coupled state $|2\rangle$, and then perform a transformation to the basis comprised of the dark and bright states, while we apply optimal control to this transformed system. Then, we confirm the optimality of the optical pumping scheme for the original closed $\Lambda$-system using numerical optimal control. We also show numerically that optical pumping remains optimal when the decay rate to the target state is larger than that to the initial state or the two rates are not very different from each other. The present work is expected to find application in various tasks of quantum information processing, where such systems are involved.

The paper has the following structure. In Sec. \ref{sec:system} we present the closed $\Lambda$-system and for large decay rates we perform the adiabatic elimination of the excited intermediate state and the transformation to the dark-bright basis. In Sec. \ref{sec:optimal} we apply optimal control theory to the transformed system and prove the optimality of optical pumping, while in Sec. \ref{sec:numerical} we confirm this theoretical result for the original system using numerical optimal control. Sec. \ref{sec:con} concludes this work.

\section{Closed $\Lambda$-system with large decay rates of the intermediate excited state}

\label{sec:system}

\begin{figure}
\centering
\includegraphics[width=.7\linewidth]{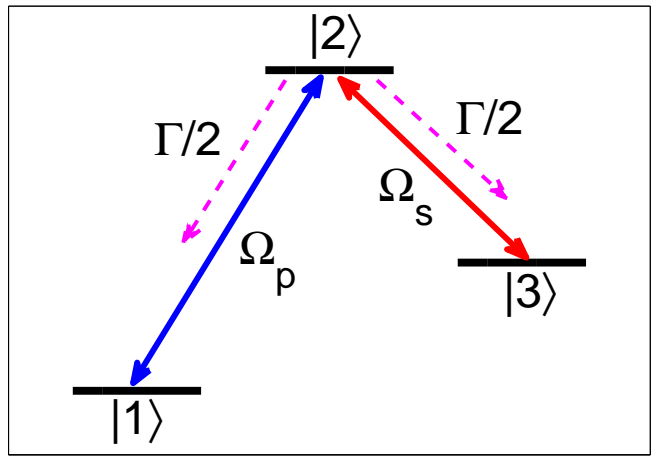}
\caption{A closed $\Lambda$-system where the intermediate excited state $|2\rangle$ decays with equal rates $\Gamma/2$ to states $|1\rangle$ and $|3\rangle$.}
\label{fig:Lambda}
\end{figure}

The coherent interaction between the $\Lambda$-system shown in Fig. \ref{fig:Lambda} and the applied electromagnetic fields is given by the Hamiltonian
\begin{equation}
\label{Hamiltonian}
H=
\frac{\hbar}{2}
\left(\begin{array}{ccc}
    0 & \Omega_p & 0 \\
    \Omega_p & 2\Delta & \Omega_s \\
    0 & \Omega_s & 2\delta
  \end{array}\right),
\end{equation}
where $\Omega_p(t), \Omega_s(t)$ denote the Rabi frequencies of the pump and Stokes fields, respectively, while $\Delta,\delta$ are the one- and two-photon detunings, which for simplicity we set to zero, $\Delta=\delta=0$. We impose a bound on the field amplitudes through the constraint
\begin{equation}
\label{constraint}
\Omega^2_p(t)+\Omega^2_s(t)=\Omega_0^2,
\end{equation}
thus
\begin{equation}
\label{angle}
\Omega_p(t)=\Omega_0\sin{\theta},\quad \Omega_s(t)=\Omega_0\cos{\theta}, \quad \tan{\theta}=\frac{\Omega_p(t)}{\Omega_s(t)}.
\end{equation}
The relaxation process is schematically displayed in Fig. \ref{fig:Lambda}, where the excited state $|2\rangle$ decays with equal rates $\Gamma/2$ to states $|1\rangle, |3\rangle$. This is the case when the lower states $|1\rangle, |3\rangle$ are the magnetic sublevels $m=-1, m=1$ of a $J=1$ level and the excited state $|2\rangle$ is the $m=0$ sublevel of a $J=0$ or $J=1$ level \cite{Vitanov01}. The density matrix of the $\Lambda$-system evolves according to the equation
\begin{equation}
\label{evolution}
i\hbar\dot{\rho}=[H,\rho]+D(\rho),
\end{equation}
where the matrix
\begin{equation}
\label{dissipator}
D(\rho)=
-i\hbar\frac{\Gamma}{2}
\left(\begin{array}{ccc}
    -\rho_{22} & \rho_{12} & 0 \\
    \rho_{21} & 2\rho_{22} & \rho_{23} \\
    0 & \rho_{32} & -\rho_{22}
  \end{array}\right)
\end{equation}
models relaxation \cite{Ivanov05}.

For large decay rates $\Gamma$, the matrix elements involving state $|2\rangle$ quickly reach a steady value, and we can eliminate them adiabatically from the equations. In order to see explicitly how this is done, we write down the equations for the matrix elements of $\rho$ from Eq. (\ref{evolution})
\begin{numparts}
\begin{eqnarray}
\dot{\rho}_{11} &= \frac{\Omega_p}{2i}(\rho_{21}-\rho_{12})+\frac{\Gamma}{2}\rho_{22},\label{dr11}\\
\dot{\rho}_{22} &= \frac{1}{2i}[\Omega_p(\rho_{12}-\rho_{21})+\Omega_s(\rho_{32}-\rho_{23})]-\Gamma\rho_{22},\label{dr22}\\
\dot{\rho}_{33} &= \frac{\Omega_s}{2i}(\rho_{23}-\rho_{32})+\frac{\Gamma}{2}\rho_{22},\label{dr33}\\
\dot{\rho}_{12} &= \frac{1}{2i}[\Omega_p(\rho_{22}-\rho_{11})-\Omega_s\rho_{13}]-\frac{\Gamma}{2}\rho_{12},\label{dr12}\\
\dot{\rho}_{23} &= \frac{1}{2i}[\Omega_p\rho_{13}+\Omega_s(\rho_{33}-\rho_{22})]-\frac{\Gamma}{2}\rho_{23},\label{dr23}\\
\dot{\rho}_{13} &= \frac{1}{2i}[\Omega_p\rho_{23}-\Omega_s\rho_{12}].\label{dr13}
\end{eqnarray}
\end{numparts}
By setting $\dot{\rho}_{22}=\dot{\rho}_{12}=\dot{\rho}_{23}=0$ we obtain
\begin{numparts}
\label{eliminated}
\begin{eqnarray}
\rho_{22} &= \frac{1}{2i\Gamma}[\Omega_p(\rho_{12}-\rho_{21})+\Omega_s(\rho_{32}-\rho_{23})],\label{r22}\\
\rho_{12} &= \frac{1}{i\Gamma}[\Omega_p(\rho_{22}-\rho_{11})-\Omega_s\rho_{13}]\approx -\frac{1}{i\Gamma}(\Omega_p\rho_{11}+\Omega_s\rho_{13}),\label{r12}\\
\rho_{23} &= \frac{1}{i\Gamma}[\Omega_p\rho_{13}+\Omega_s(\rho_{33}-\rho_{22})]\approx \frac{1}{i\Gamma}(\Omega_p\rho_{13}+\Omega_s\rho_{33}),\label{r23}
\end{eqnarray}
\end{numparts}
where the approximations in Eqs. (\ref{r12}), (\ref{r23}) are obtained by omitting the small terms involving $\rho_{22}$. Although $\rho_{22}$ is small compared to the matrix elements involving states $|1\rangle, |3\rangle$ only, it is still necessary to find it, since in Eqs. (\ref{dr11}), (\ref{dr33}) is multiplied by the large decay rate $\Gamma$, thus the term $\Gamma\rho_{22}$ cannot be omitted. If we plug Eqs. (\ref{r12}), (\ref{r23}) into Eq. (\ref{r22}) we get
\begin{equation}
\label{r22_f}
\rho_{22}=\frac{1}{\Gamma^2}[\Omega_p^2\rho_{11}+\Omega_s^2\rho_{33}+\Omega_p\Omega_s(\rho_{13}+\rho_{31})].
\end{equation}
If we substitute Eqs. (\ref{r12}), (\ref{r23}), (\ref{r22_f}) in Eqs. (\ref{dr11}), (\ref{dr33}), (\ref{dr13}) we obtain
\begin{numparts}
\label{adiabatic_elim}
\begin{eqnarray}
\dot{\rho}_{11} &= \frac{\Omega^2_0}{2\Gamma}(-\sin^2{\theta}\rho_{11}+\cos^2{\theta}\rho_{33}),\label{elim11}\\
\dot{\rho}_{13} &= -\frac{\Omega^2_0}{2\Gamma}[\rho_{13}+\sin{\theta}\cos{\theta}(\rho_{11}+\rho_{33})],\label{elim13}\\
\dot{\rho}_{33} &= \frac{\Omega^2_0}{2\Gamma}(\sin^2{\theta}\rho_{11}-\cos^2{\theta}\rho_{33}),\label{elim33}
\end{eqnarray}
\end{numparts}
where we have also used Eq. (\ref{angle}) for $\Omega_p, \Omega_s$.

The next step is to transform the equations to the basis of the dark and bright states, defined in the usual way
\begin{numparts}
\label{db_basis}
\begin{eqnarray}
|b(t)\rangle &= \sin{\theta(t)}|1\rangle+\cos{\theta(t)}|3\rangle, \label{bright}\\
|d(t)\rangle &= \cos{\theta(t)}|1\rangle-\sin{\theta(t)}|3\rangle. \label{dark}
\end{eqnarray}
\end{numparts}
The transformation connecting the density matrices
\begin{equation}
\label{matrices}
\rho'=
\left(\begin{array}{cc}
    \rho_{11} & \rho_{13} \\ \rho_{31} & \rho_{33}
  \end{array}\right),
\quad
\tilde{\rho}=
\left(\begin{array}{cc}
    \rho_{dd} & \rho_{db} \\ \rho_{bd} & \rho_{bb}
  \end{array}\right)
\end{equation}
is
\begin{equation}
\label{transformation}
\tilde{\rho}=R\rho' R^{-1},
\quad
R=
\left(\begin{array}{cc}
    \cos{\theta} & -\sin{\theta} \\ \sin{\theta} & \cos{\theta}
  \end{array}\right).
\end{equation}
Using this equation along with Eqs. (\ref{elim11})-(\ref{elim33}), we find that the population difference between the bright and dark states and the real part of the corresponding coherence,
\begin{numparts}
\begin{eqnarray}
x &= \rho_{bb}-\rho_{dd},\label{x_def}\\
y &= 2\mbox{Re}(\rho_{db}),\label{y_def}
\end{eqnarray}
\end{numparts}
satisfy the equations
\begin{numparts}
\label{system1}
\begin{eqnarray}
\dot{y} &=-2\dot{\theta}x-\frac{\Omega_0^2}{2\Gamma}y \label{y1}\\
\dot{x} & =-\frac{\Omega_0^2}{2\Gamma}(x+1)+2\dot{\theta}y. \label{x1}
\end{eqnarray}
\end{numparts}
Note that these are the same equations derived in Ref. \cite{Ivanov05} by first making the transformation to the dark-bright basis and then performing the adiabatic elimination.
If we normalize time as $dt'=(\Omega_0^2dt)/(2\Gamma)$ then we obtain the control system
\begin{numparts}
\label{system}
\begin{eqnarray}
\dot{y} &=-2ux-y \label{y}\\
\dot{x} &=-(x+1)+2uy, \label{x}\\
\dot{\theta}&=u, \label{theta}
\end{eqnarray}
\end{numparts}
where we consider as control function the time derivative $u$ of the angle $\theta$.
For a fixed duration $T$ and starting from $x(0)=-1, y(0)=0$, we would like to find the control $u(t)$, $0\leq t \leq T$, which minimizes the final value $x(T)$ while changing angle $\theta$ from $\theta(0)=0$ to $\theta(T)=\pi/2$. By minimizing $x(T)$ we actually maximize $\rho_{dd}(T)$, see Eq. (\ref{x_def}) and note that $\rho_{dd}+\rho_{bb}=1$ within the approximation of adiabatic elimination. For $\theta(T)=\pi/2$ this is equivalent to maximizing the final population $\rho_{33}(T)$.

\section{Optimality of the optical pumping scheme}

\label{sec:optimal}

In order to solve the optimal control problem defined in the previous section, we need to formulate the control Hamiltonian \cite{Bryson} corresponding to system (\ref{y})-(\ref{theta}). This is a mathematical construction whose maximization results in the optimization of the target quantity, here $x(T)$. It is formed by adjoining to each state equation a conjugate variable (Lagrange multiplier) as follows
\begin{eqnarray}
\label{Hc}
H_c&=\lambda_x\dot{x}+\lambda_y\dot{y}+\mu\dot{\theta}\nonumber\\
   &=\lambda_x(2uy-x-1)+\lambda_y(-2ux-y)+\mu u \nonumber\\
   &=(2\lambda_xy-2\lambda_yx+\mu)u-\lambda_x(x+1)-\lambda_yy,
\end{eqnarray}
where $\lambda_x,\lambda_y,\mu$ are the Lagrange multipliers corresponding to the state variables $x,y,\theta$. They satisfy the adjoint equations
\begin{numparts}
\label{adjoint}
\begin{eqnarray}
\dot{\lambda_y} &=-\frac{\partial H_c}{\partial y}=-2u\lambda_x+\lambda_y, \label{ly}\\
\dot{\lambda_x} &=-\frac{\partial H_c}{\partial x}=2u\lambda_y+\lambda_x, \label{lx}
\end{eqnarray}
\end{numparts}
while $\mu$ is constant since $\theta$ is a cyclic variable.
Due to the construction of $H_c$ the state equations can also be expressed as
\begin{numparts}
\begin{eqnarray}
\dot{y}&=\frac{\partial H_c}{\partial \lambda_y}&=-2ux-y,\\
\dot{x}&=\frac{\partial H_c}{\partial \lambda_x}&=-(x+1)+2uy,\\
\dot{\theta}&=\frac{\partial H_c}{\partial \mu}&=u,
\end{eqnarray}
\end{numparts}
which justifies the term Hamiltonian for $H_c$.

According to the principles of optimal control theory, the optimal $u(t)$ is chosen to maximize the control Hamiltonian $H_c$ \cite{Bryson}. Note that we have not imposed any bound on $u$, thus even infinite values are allowed momentarily, corresponding to instantaneous jumps in the angle $\theta$. Since $H_c$ is a linear function of $u$ with coefficient $\phi=2\lambda_xy-2\lambda_yx+\mu$, if $\phi\neq 0$ for a finite interval then the corresponding optimal control should be $\pm\infty$ for the whole interval, which is obviously unphysical. We conclude that $\phi=0$ almost everywhere, except some isolated points where jumps in the angle $\theta$ can occur. The optimal control which maintains the condition $\phi=0$ is called \emph{singular} \cite{Bryson}, while the delta pulses which can modify instantaneously the angle $\theta$ by a finite amount are called \emph{bang} controls. In order to find the singular optimal control $u_s$ we additionally use the conditions $\dot{\phi}=\ddot{\phi}=0$, which also hold on a singular arc since it is already $\phi=0$, and obtain the following equations
\begin{numparts}
\label{switching}
\begin{eqnarray}
\phi &=2\lambda_xy-2\lambda_yx+\mu=0,\label{phi}\\
\dot{\phi} &=2\lambda_y=0,\label{dphi}\\
\ddot{\phi} &=-4u\lambda_x+2\lambda_y=0.\label{ddphi}
\end{eqnarray}
\end{numparts}
From Eq. (\ref{dphi}) we get $\lambda_y=0$ and, using this in Eq. (\ref{ddphi}), we find $u_s=0$ or $\lambda_x=0$. The choice $\lambda_x=0$ along with the relation $\lambda_y=0$, when used in Eq. (\ref{phi}), lead also to $\mu=0$ which is not allowed since, according to optimal control theory, the adjoint variables cannot be simultaneously zero. Consequently, the singular optimal control is
\begin{equation}
\label{singular_optimal}
u_s=0,
\end{equation}
which implies that on the singular arc the angle $\theta$ remains constant.

A pulse-sequence should start with a bang control since for $u=0$ the starting point $(x_0, y_0, \theta_0)=(-1, 0, 0)$ is an equilibrium point of system (\ref{y})-(\ref{theta}), and in principle it may contain more bang controls, separated by finite time intervals of singular control. Nevertheless, we will prove that the optimal pulse-sequence has the bang-singular form, with a delta pulse $u(t)=(\pi/2)\delta(t)$ which instantaneously changes $\theta$ from $0$ to $\pi/2$, followed by the application of singular control $u_s=0$ for the whole time interval $T$. Since $\Omega_p(t)=\Omega_0\sin{\theta(t)}, \Omega_s(t)=\Omega_0\cos{\theta(t)}$, this corresponds to $\Omega_p=\Omega_0, \Omega_s=0$ for the whole duration $T$, which is actually the method of optical pumping using an orthogonal pulse. In order to prove the optimality of optical pumping, we consider a general pulse-sequence of the form bang-singular-bang-singular-...bang-singular, composed by $n$ bang pulses with strengths $\theta_i, i=1,2,\ldots,n$, satisfying $\sum_{i=1}^n\theta_i=\pi/2$ so $\theta(T)=\pi/2$, each of them followed by an interval $t_i$ of singular control, so $\sum_{i=1}^nt_i=T$. We will find $X_n$, the final value of $x(T)$ under the application of the previously described $n$-bang pulse-sequence, and show that $X_n\geq X_1$.

Let $(x'_i,y'_i)$ be the values of $(x,y)$ after the application of $i$-th bang pulse with strength $\theta_i$, and $(x_i,y_i)$ the values of $(x,y)$ after the application of the subsequent singular control of duration $t_i$. From Eqs. (\ref{y}), (\ref{x}) we see that the application of a delta pulse with strength $\theta_i$ corresponds to a clockwise rotation by an angle $2\int u(t)dt=2\theta_i$ on the $xy$-plane, thus
\begin{equation}
\label{bang}
\left(\begin{array}{c}
    x'_i\\
    y'_i
\end{array}\right)
=
\left(\begin{array}{cc}
    \cos{2\theta_i} & \sin{2\theta_i} \\ -\sin{2\theta_i} & \cos{2\theta_i}
  \end{array}\right)
\left(\begin{array}{c}
    x_{i-1}\\
    y_{i-1}
\end{array}\right).
\end{equation}
On the following singular arc of duration $t_i$ it is $u(t)=0$, thus
\begin{equation}
\label{singular}
\left(\begin{array}{c}
    x_i\\
    y_i
\end{array}\right)
=
e^{-t_i}
\left(\begin{array}{c}
    x'_i\\
    y'_i
\end{array}\right)
+
\left(\begin{array}{c}
    e^{-t_i}-1\\
    0
\end{array}\right).
\end{equation}
Combining these equations we find the inductive relation
\begin{equation}
\label{induction}
\left(\begin{array}{c}
    x_i\\
    y_i
\end{array}\right)
=
e^{-t_i}
\left(\begin{array}{cc}
    \cos{2\theta_i} & \sin{2\theta_i} \\ -\sin{2\theta_i} & \cos{2\theta_i}
  \end{array}\right)
\left(\begin{array}{c}
    x_{i-1}\\
    y_{i-1}
\end{array}\right)
+
\left(\begin{array}{c}
    e^{-t_i}-1\\
    0
\end{array}\right).
\end{equation}
For $i=1$ and since $(x_0, y_0)=(-1, 0)$ we obtain
\begin{numparts}
\begin{eqnarray}
x_1 &= e^{-t_1}\left(1-\cos{2\theta_1}\right)-1,\label{x_1}\\
y_1 &= e^{-t_1}\sin{2\theta_1}.\label{y_1}
\end{eqnarray}
\end{numparts}
Note that for $n=1$ (one bang pulse, optical pumping), it is $\theta_1=\pi/2$ and $t_1=T$, thus
\begin{equation}
\label{X1}
X_1=x(T)=x_1=2e^{-T}-1.
\end{equation}
For $i=2$, using Eqs. (\ref{induction}) and (\ref{x_1}), (\ref{y_1})  we find
\begin{numparts}
\begin{eqnarray}
x_2 &= e^{-(t_1+t_2)}\left[\cos{2\theta_2}-\cos{2(\theta_1+\theta_2)}\right]\nonumber\\
    &+ e^{-t_2}\left(1-\cos{2\theta_2}\right)-1,\label{x_2}\\
y_2 &= e^{-(t_1+t_2)}\left[\sin{2(\theta_1+\theta_2)}-\sin{2\theta_2}\right]\nonumber\\
    &+ e^{-t_2}\sin{2\theta_2}.\label{y_2}
\end{eqnarray}
\end{numparts}
Analogously, for $i=3$ we obtain
\begin{numparts}
\begin{eqnarray}
x_3 &= e^{-(t_1+t_2+t_3)}\left[\cos{2(\theta_2+\theta_3)}-\cos{2(\theta_1+\theta_2+\theta_3)}\right]\nonumber\\
    &+ e^{-(t_2+t_3)}\left[\cos{2\theta_3}-\cos{2(\theta_2+\theta_3)}\right]\nonumber\\
    &+ e^{-t_3}\left(1-\cos{2\theta_3}\right)-1,\label{x_3}\\
y_3 &= e^{-(t_1+t_2+t_3)}\left[\sin{2(\theta_1+\theta_2+\theta_3)}-\sin{2(\theta_2+\theta_3)}\right]\nonumber\\
    &+ e^{-(t_2+t_3)}\left[\sin{2(\theta_2+\theta_3)}-\sin{2\theta_3}\right]\nonumber\\
    &+ e^{-t_3}\sin{2\theta_3}.\label{y_3}
\end{eqnarray}
\end{numparts}
By inspecting the above equations we conclude to the following pattern
\begin{numparts}
\label{xyn}
\begin{eqnarray}
x_n &= e^{-(t_1+t_2+\ldots+t_n)}\left[\cos{2(\theta_2+\theta_3+\ldots+\theta_n)}-\cos{2(\theta_1+\theta_2+\ldots+\theta_n)}\right]\nonumber\\
    &+ e^{-(t_2+\ldots+t_n)}\left[\cos{2(\theta_3+\ldots+\theta_n)}-\cos{2(\theta_2+\theta_3+\ldots+\theta_n)}\right]\nonumber\\
    &+\ldots\nonumber\\
    &+ e^{-t_n}\left(1-\cos{2\theta_n}\right)-1,\label{x_n}\\
y_n &= e^{-(t_1+t_2+\ldots+t_n)}\left[\sin{2(\theta_1+\theta_2+\ldots+\theta_n)}-\sin{2(\theta_2+\theta_3+\ldots+\theta_n)}\right]\nonumber\\
    &+ e^{-(t_2+\ldots+t_n)}\left[\sin{2(\theta_2+\theta_3+\ldots+\theta_n)}-\sin{2(\theta_3+\ldots+\theta_n)}\right]\nonumber\\
    &+\ldots\nonumber\\
    &+ e^{-t_n}\sin{2\theta_n},\label{y_n}
\end{eqnarray}
\end{numparts}
which we subsequently prove using induction.

We assume that Eqs. (\ref{x_n}), (\ref{y_n}) hold for some integer $n$ and along with Eq. (\ref{induction}) we use them to find $x_{n+1}$:
\begin{eqnarray}
\label{step1}
\fl x_{n+1} &= e^{-t_{n+1}}(\cos{2\theta_{n+1}}x_n+\sin{2\theta_{n+1}}y_n)+e^{-t_{n+1}}-1\nonumber\\
\fl      &=e^{-(t_1+t_2+\ldots+t_n+t_{n+1})}\left[\cos{2\theta_{n+1}}\cos{2(\theta_2+\theta_3+\ldots+\theta_n)}-\cos{2\theta_{n+1}}\cos{2(\theta_1+\theta_2+\ldots+\theta_n)}\right]\nonumber\\
\fl        &+ e^{-(t_2+\ldots+t_n+t_{n+1})}\left[\cos{2\theta_{n+1}}\cos{2(\theta_3+\ldots+\theta_n)}-\cos{2\theta_{n+1}}\cos{2(\theta_2+\theta_3+\ldots+\theta_n)}\right]\nonumber\\
\fl        &+\ldots\nonumber\\
\fl        &+ e^{-(t_n+t_{n+1})}\left(\cos{2\theta_{n+1}}-\cos{2\theta_{n+1}}\cos{2\theta_n}\right)-e^{-t_{n+1}}\cos{2\theta_{n+1}}\nonumber\\
\fl &+ e^{-(t_1+t_2+\ldots+t_n+t_{n+1})}\left[\sin{2\theta_{n+1}}\sin{2(\theta_1+\theta_2+\ldots+\theta_n)}-\sin{2\theta_{n+1}}\sin{2(\theta_2+\theta_3+\ldots+\theta_n)}\right]\nonumber\\
\fl    &+ e^{-(t_2+\ldots+t_n+t_{n+1})}\left[\sin{2\theta_{n+1}}\sin{2(\theta_2+\theta_3+\ldots+\theta_n)}-\sin{2\theta_{n+1}}\sin{2(\theta_3+\ldots+\theta_n)}\right]\nonumber\\
\fl    &+\ldots\nonumber\\
\fl    &+ e^{-(t_n+t_{n+1})}\sin{2\theta_{n+1}}\sin{2\theta_n}\nonumber\\
\fl    &+e^{-t_{n+1}}-1.
\end{eqnarray}
Combining and rearranging the terms which multiply the same exponentials we obtain
\begin{eqnarray}
\label{step2}
\fl x_{n+1}      &=e^{-(t_1+t_2+\ldots+t_n+t_{n+1})}[\cos{2\theta_{n+1}}\cos{2(\theta_2+\theta_3+\ldots+\theta_n)}-\sin{2\theta_{n+1}}\sin{2(\theta_2+\theta_3+\ldots+\theta_n)}\nonumber\\
\fl      &-\cos{2\theta_{n+1}}\cos{2(\theta_1+\theta_2+\ldots+\theta_n)}+\sin{2\theta_{n+1}}\sin{2(\theta_1+\theta_2+\ldots+\theta_n)}]\nonumber\\
\fl      &+ e^{-(t_2+\ldots+t_n+t_{n+1})}[\cos{2\theta_{n+1}}\cos{2(\theta_3+\ldots+\theta_n)}-\sin{2\theta_{n+1}}\sin{2(\theta_3+\ldots+\theta_n)}\nonumber\\
\fl      &-\cos{2\theta_{n+1}}\cos{2(\theta_2+\theta_3+\ldots+\theta_n)}+\sin{2\theta_{n+1}}\sin{2(\theta_2+\theta_3+\ldots+\theta_n)}]\nonumber\\
\fl      &+\ldots\nonumber\\
\fl      &+ e^{-(t_n+t_{n+1})}\left(\cos{2\theta_{n+1}}-\cos{2\theta_{n+1}}\cos{2\theta_n}+\sin{2\theta_{n+1}}\sin{2\theta_n}\right)\nonumber\\
\fl    &+e^{-t_{n+1}}(1-\cos{2\theta_{n+1}})-1.
\end{eqnarray}
Using the trigonometric identity $\cos{(\alpha+\beta)}=\cos{\alpha}\cos{\beta}-\sin{\alpha}\sin{\beta}$ for the terms multiplying each exponential we finally get
\begin{eqnarray}
\label{step3}
\fl x_{n+1} &= e^{-(t_1+t_2+\ldots+t_n+t_{n+1})}\left[\cos{2(\theta_2+\theta_3+\ldots+\theta_n+\theta_{n+1})}-\cos{2(\theta_1+\theta_2+\ldots+\theta_n+\theta_{n+1})}\right]\nonumber\\
\fl    &+ e^{-(t_2+\ldots+t_n+t_{n+1})}\left[\cos{2(\theta_3+\ldots+\theta_n+\theta_{n+1})}-\cos{2(\theta_2+\theta_3+\ldots+\theta_n+\theta_{n+1})}\right]\nonumber\\
\fl    &+\ldots\nonumber\\
\fl    &+ e^{-(t_n+t_{n+1})}\left[\cos{2\theta_{n+1}}-\cos{2(\theta_n+\theta_{n+1})}\right]\nonumber\\
\fl    &+ e^{-t_{n+1}}\left(1-\cos{2\theta_{n+1}}\right)-1,
\end{eqnarray}
which proves the induction step and thus the validity of Eq. (\ref{x_n}). Working analogously we can prove Eq. (\ref{y_n}).

Now for a pulse sequence of the form bang-singular-bang-singular-...bang-singular considered above with $n$ bangs, the final value $x(T)$ is $X_n=x(T)=x_n$. By rearranging the terms in Eq. (\ref{x_n}) we get
\begin{eqnarray}
X_n &= -e^{-(t_1+t_2+\ldots+t_n)}\cos{2(\theta_1+\theta_2+\ldots+\theta_n)}\nonumber\\
    &+ [e^{-(t_1+t_2+\ldots+t_n)}-e^{-(t_2+\ldots+t_n)}]\cos{2(\theta_2+\theta_3+\ldots+\theta_n)}\nonumber\\
    &+ [e^{-(t_2+t_3+\ldots+t_n)}-e^{-(t_3+t_4+\ldots+t_n)}]\cos{2(\theta_3+\theta_4+\ldots+\theta_n)}\nonumber\\
    &+\ldots\nonumber\\
    &+ (e^{-(t_{n-1}+t_{n})}-e^{-t_n})\cos{2\theta_n}\nonumber\\
    &+ e^{-t_n}-1.\label{rearrange}
\end{eqnarray}
Observe that, since $\sum_{i=1}^n\theta_i=\pi/2$ and $\sum_{i=1}^nt_i=T$, the term in the first line is fixed to $e^{-T}$. In the subsequent lines, except the last one, the difference of exponentials multiplying each cosine is negative, since the positive left exponential has always an extra negative term in the exponent than the negative right exponential. Each of these lines is minimized when the corresponding cosine attains its maximum value one, which leads to the conditions $\theta_2+\theta_3+\ldots+\theta_n=0$, $\theta_3+\theta_4+\ldots+\theta_n=0$, ..., $\theta_n=0$. From these relations we obtain $\theta_1=\pi/2$ and $\theta_2=\theta_3=\ldots=\theta_n=0$, while
\begin{eqnarray}
X_n &\geq e^{-(t_1+t_2+\ldots+t_n)}\nonumber\\
    &+ e^{-(t_1+t_2+\ldots+t_n)}-\cancel{e^{-(t_2+t_3+\ldots+t_n)}}\nonumber\\
    &+ \cancel{e^{-(t_2+t_3+\ldots+t_n)}}-\cancel{e^{-(t_3+t_4+\ldots+t_n)}}\nonumber\\
    &+\ldots\nonumber\\
    &+ \cancel{e^{-(t_{n-1}+t_{n})}}-\cancel{e^{-t_n}}\nonumber\\
    &+ \cancel{e^{-t_n}}-1\nonumber\\
    &=2e^{-T}-1\nonumber\\
    &=X_1.\label{minimum}
\end{eqnarray}
We have thus proved the optimality of the bang-singular pulse-sequence, corresponding to optical pumping. Note that in this case we have $\rho_{bb}(T)-\rho_{dd}(T)=X_1$, thus $1-2\rho_{dd}(T)=2e^{-T}-1$ and $\rho_{33}(T)=\rho_{dd}(T)=1-e^{-T}$. If we restore the dimensionality of time we find
\begin{equation}
\label{opt_pump_eff}
\rho_{33}(T)=1-e^{-\frac{\Omega_0^2}{2\Gamma}T}.
\end{equation}

We close this section with some observations regarding the use of bang pulses under the adiabatic approximation. First, observe from Eqs. (\ref{y})-(\ref{theta}) that bang controls implement jumps in the angle $\theta$ and instantaneous rotations of the variables $x$ and $y$. But $x, y$ express the state of the system in the dark-bright basis and from Eqs. (\ref{bright}), (\ref{dark}) we observe that this basis also changes instantaneously. The combined changes of $x, y$ and the dark-bright basis result so the state of the system, expressed in the original basis $|1\rangle, |3\rangle$, remains unchanged. This means that in Eqs. (\ref{elim11})-(\ref{elim33}), obtained with adiabatic elimination, the bang pulses appear only as instantaneous changes in $\theta$. Actually, a jump in the angle $\theta$ in incorporated in the adiabatic equations through the adiabatically eliminated elements (\ref{r22})-(\ref{r23}) in a time of the order of $1/\Gamma$, which is very small compared to the timescale $2\Gamma/\Omega_0^2$ of the adiabatic system for the case of large $\Gamma$ considered here. The second observation is that in this section we solved the mathematical optimal control problem in the more general case where bang controls are allowed, instead of restricting $\dot{\theta}$ to small values, and found the optimal solution to be a single bang pulse, which corresponds to $\Omega_p(t)=\Omega_0$ and $\Omega_s(t)=0$. This solution is actually optical pumping with a square pulse, and can be easily implemented experimentally by directly applying these values from the beginning.

\section{Verification using numerical optimal control}

\label{sec:numerical}


Here we use numerical optimization to confirm the optimality of the optical pumping scheme for large $\Gamma$, that was previously proved using adiabatic elimination, for the original system (\ref{evolution}).
From Eqs. (\ref{dr11})-(\ref{dr13}) we obtain for the real variables
$x_1=\rho_{11}, x_2=\rho_{22}, x_3=\rho_{33}, x_4=\mbox{Im}(\rho_{12}), x_5=\mbox{Im}(\rho_{23}), x_6=\mbox{Re}(\rho_{13})$ the system
\begin{numparts}
\begin{eqnarray}
\dot{x}_{1} &= -\Omega_px_4+\frac{\Gamma}{2}x_2,\label{x1}\\
\dot{x}_{2} &= \Omega_px_4-\Omega_sx_5-\Gamma x_2,\label{x2}\\
\dot{x}_{3} &= \Omega_sx_5+\frac{\Gamma}{2}x_2,\label{x3}\\
\dot{x}_{4} &= \frac{\Omega_p}{2}(x_1-x_2)+\frac{\Omega_s}{2}x_6-\frac{\Gamma}{2}x_4,\label{x4}\\
\dot{x}_{5} &= \frac{\Omega_s}{2}(x_2-x_4)-\frac{\Omega_p}{2}x_6-\frac{\Gamma}{2}x_5,\label{x5}\\
\dot{x}_{6} &= \frac{1}{2}(\Omega_px_5-\Omega_sx_4)\label{x6}
\end{eqnarray}
\end{numparts}
with initial conditions $x_1(0)=1, x_i(0)=0$, $i=2, 3,\ldots, 6$,
and for the real variables $y_1=\mbox{Re}(\rho_{12}), y_2=\mbox{Re}(\rho_{23}), y_3=\mbox{Im}(\rho_{13})$ the system
\begin{numparts}
\begin{eqnarray}
\dot{y}_{1} &= -\frac{\Omega_s}{2}y_3-\frac{\Gamma}{2}y_1,\label{y1}\\
\dot{y}_{2} &= \frac{\Omega_p}{2}y_3-\frac{\Gamma}{2}y_2,\label{y2}\\
\dot{y}_{3} &= \frac{1}{2}(\Omega_sy_1-\Omega_py_2)\label{y3}
\end{eqnarray}
\end{numparts}
with initial conditions $y_i(0)=0$, $i=1, 2, 3$. Obviously, $y_i(t)=0$ throughout, thus we concentrate on system (\ref{x1})-(\ref{x6}) with the nonzero initial conditions.

\begin{figure*}[t]
 \centering
		\begin{tabular}{cc}
     	\subfigure[$\ $]{
	            \label{fig:con_01_5}
	            \includegraphics[width=.45\linewidth]{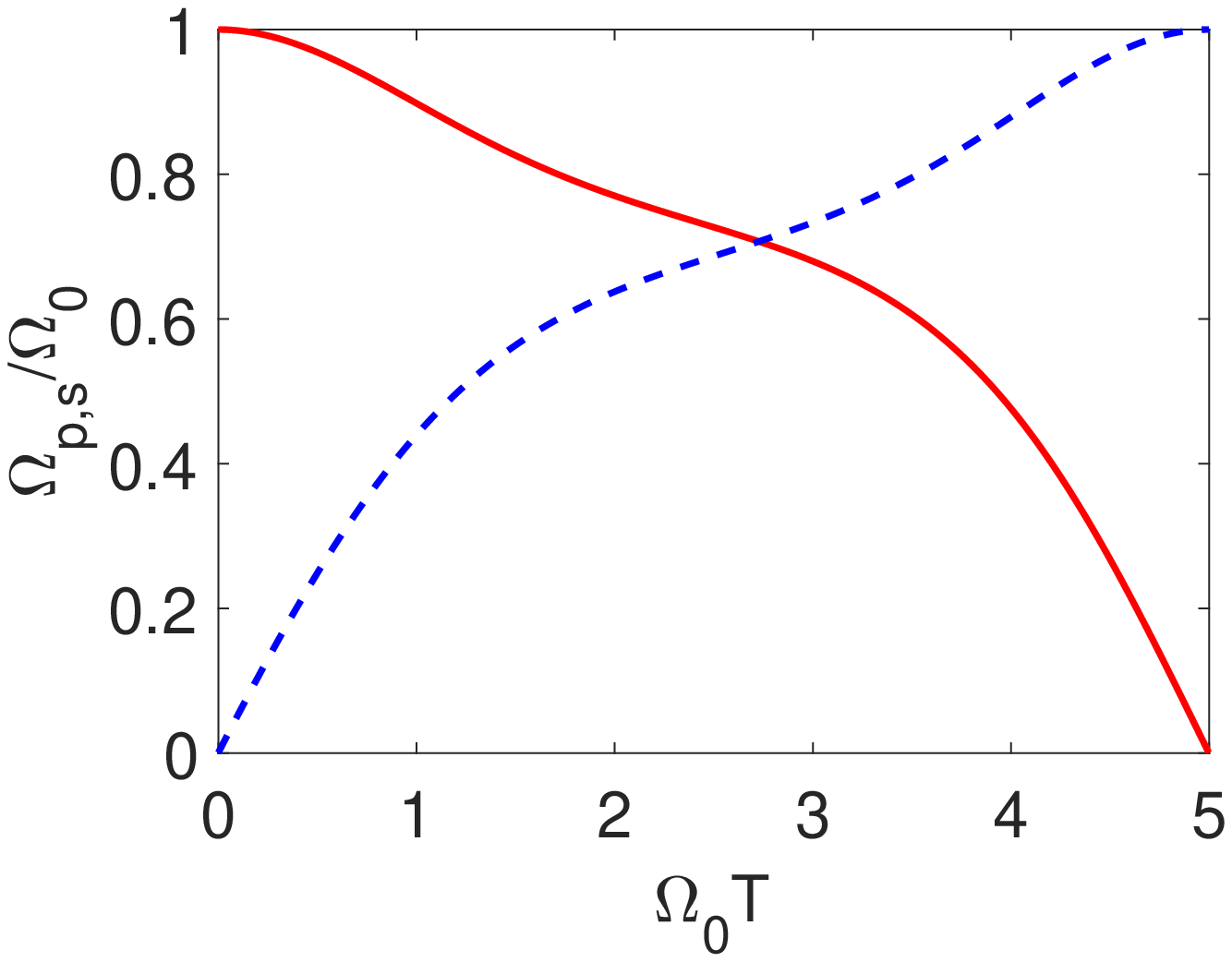}} &
       \subfigure[$\ $]{
	            \label{fig:pop_01_5}
	            \includegraphics[width=.45\linewidth]{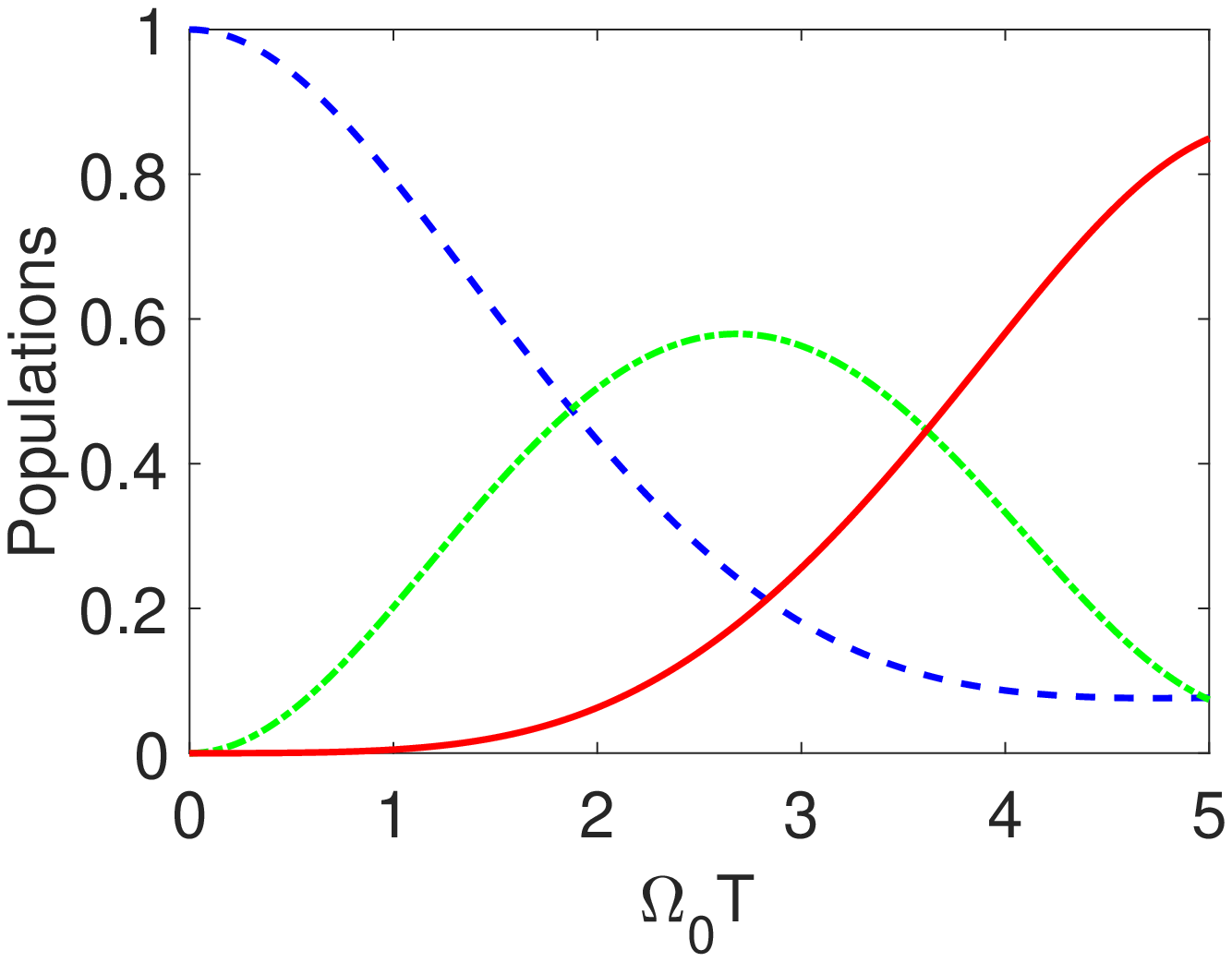}} \\
       \subfigure[$\ $]{
	            \label{fig:con_01_10}
	            \includegraphics[width=.45\linewidth]{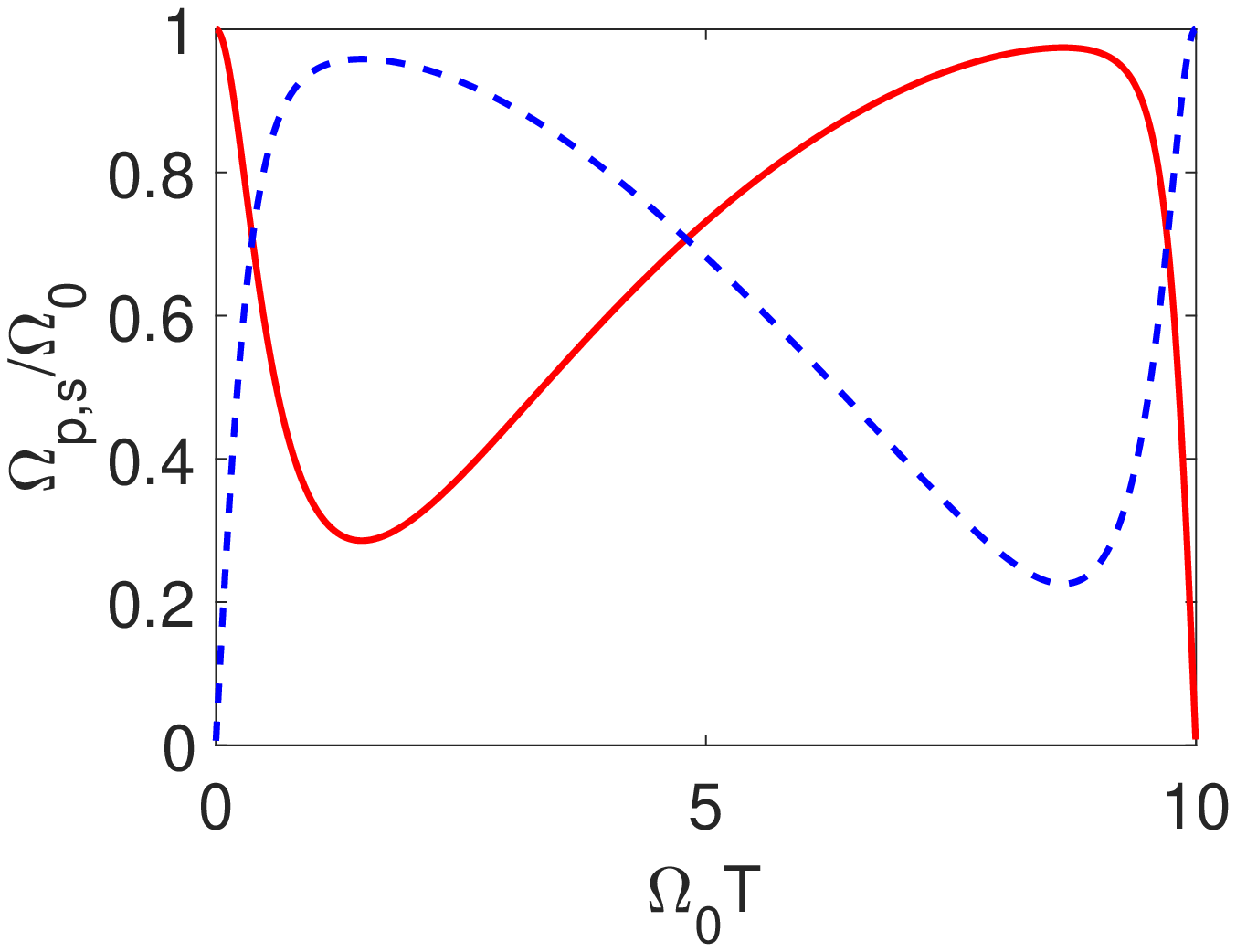}} &
       \subfigure[$\ $]{
	            \label{fig:pop_01_10}
	            \includegraphics[width=.45\linewidth]{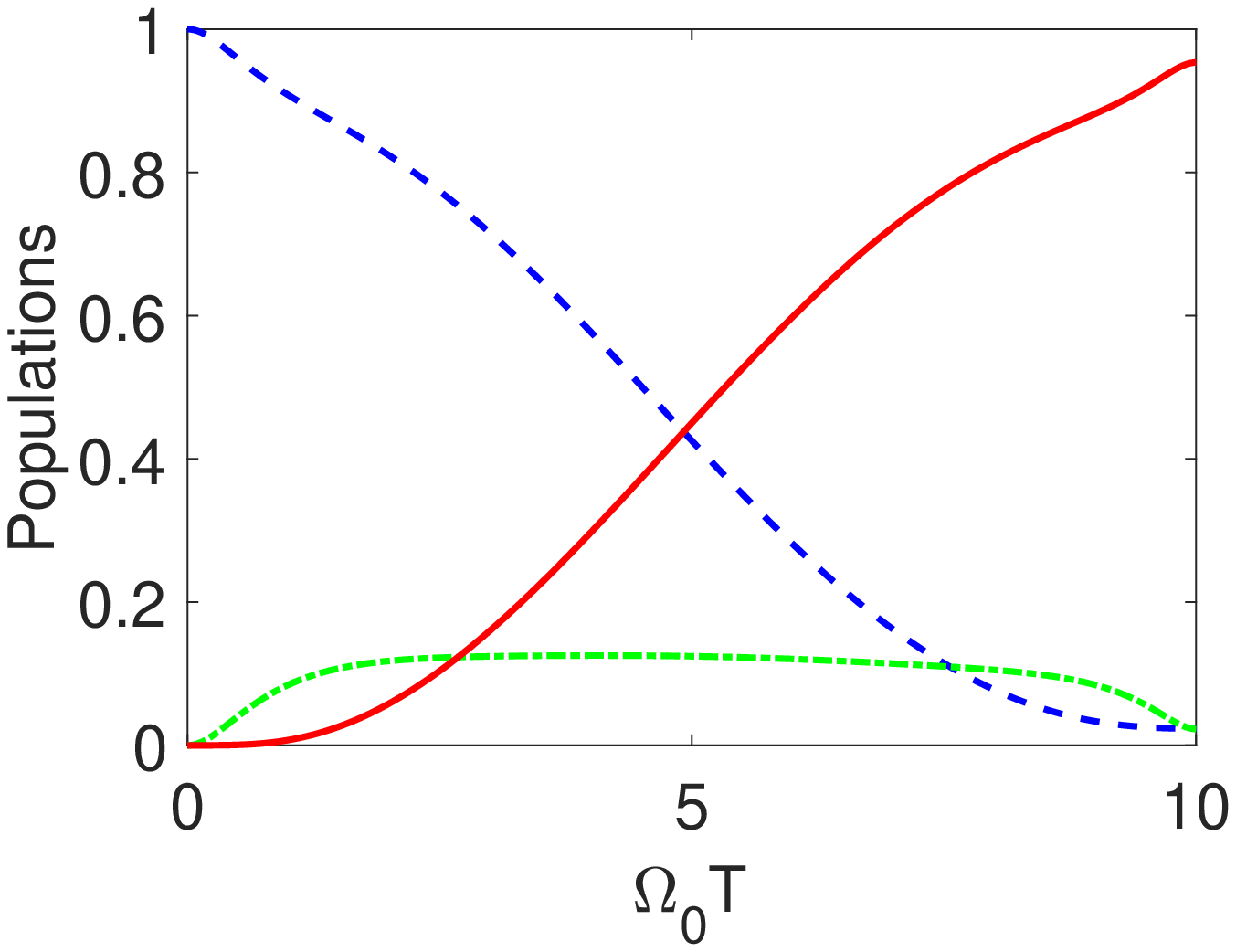}} \\
       \subfigure[$\ $]{
	            \label{fig:con_01_20}
	            \includegraphics[width=.45\linewidth]{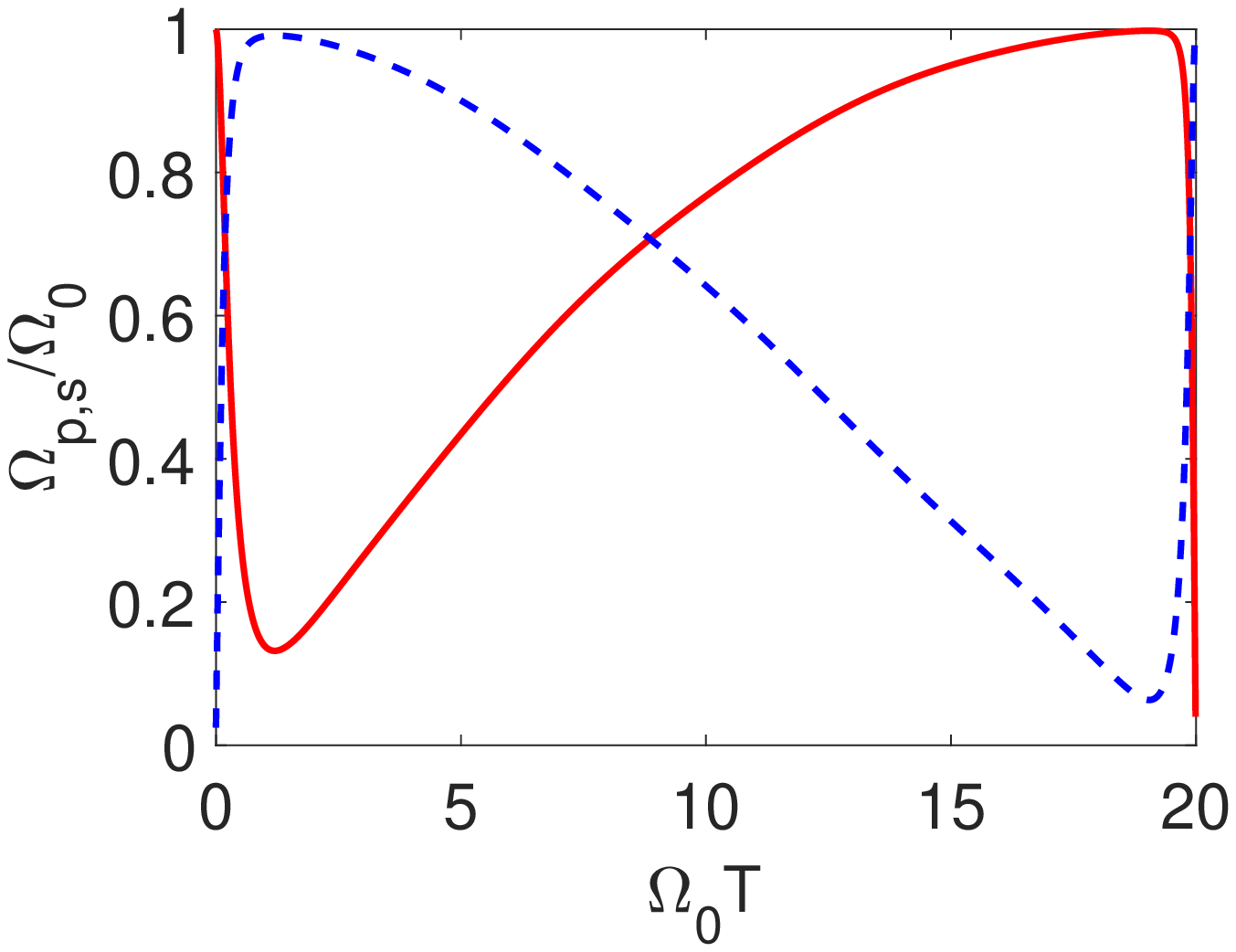}} &
       \subfigure[$\ $]{
	            \label{fig:pop_01_20}
	            \includegraphics[width=.45\linewidth]{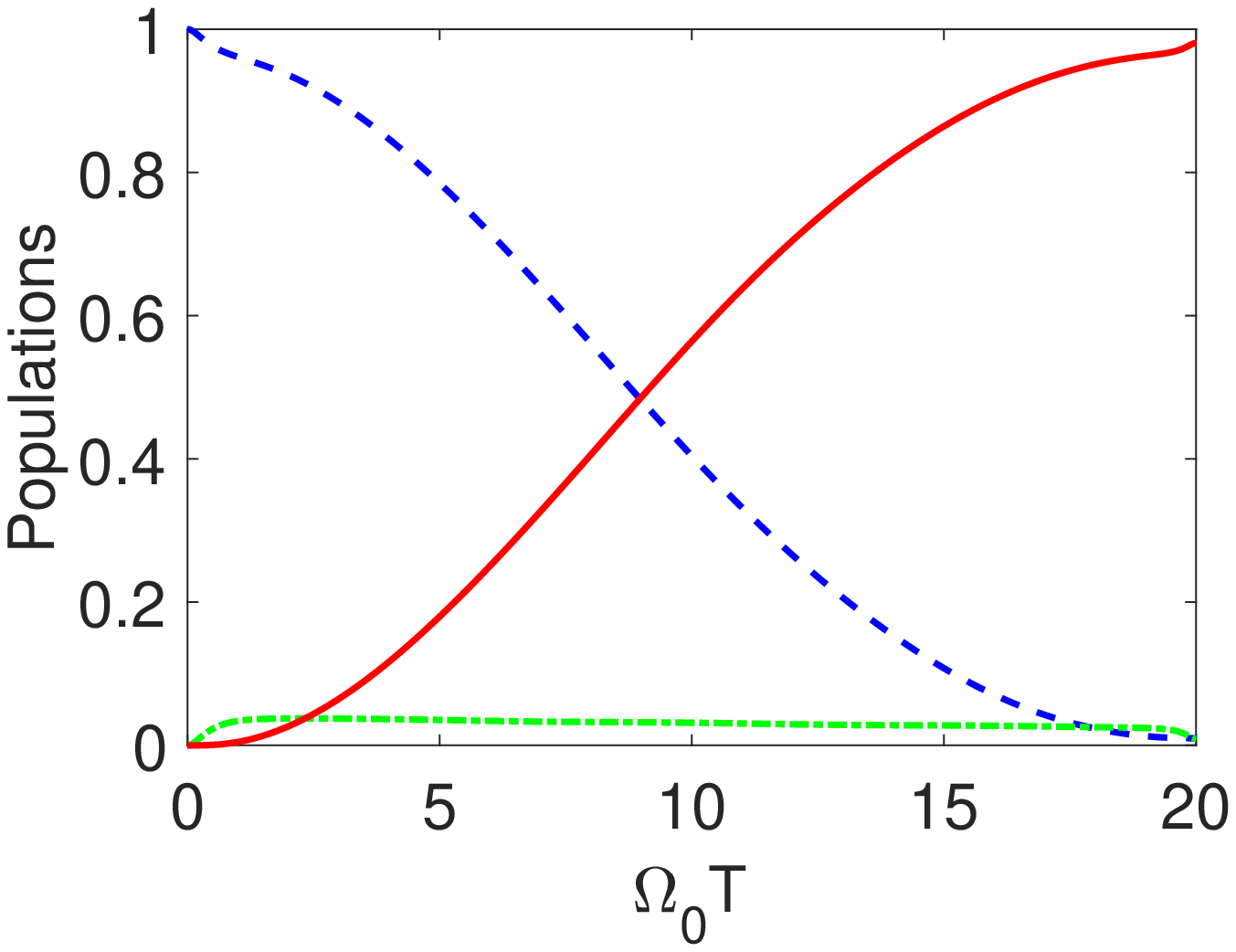}} \\
		\end{tabular}
\caption{(a, c, e) Optimal controls $\Omega_p(t)$ (red solid curve) and $\Omega_s(t)$ (blue dashed curve) obtained with numerical optimization for decay rate $\Gamma/\Omega_0=0.1$ and three different durations $\Omega_0T=5, 10, 20$ from top to bottom. (b, d, f) Corresponding evolution of populations $\rho_{11}(t)$ (blue dashed curve), $\rho_{22}(t)$ (green dashed-dotted curve), and $\rho_{33}(t)$ (red solid curve).}
\label{fig:example1}
\end{figure*}

\begin{figure*}[t]
 \centering
		\begin{tabular}{cc}
     	\subfigure[$\ $]{
	            \label{fig:con_2_10}
	            \includegraphics[width=.45\linewidth]{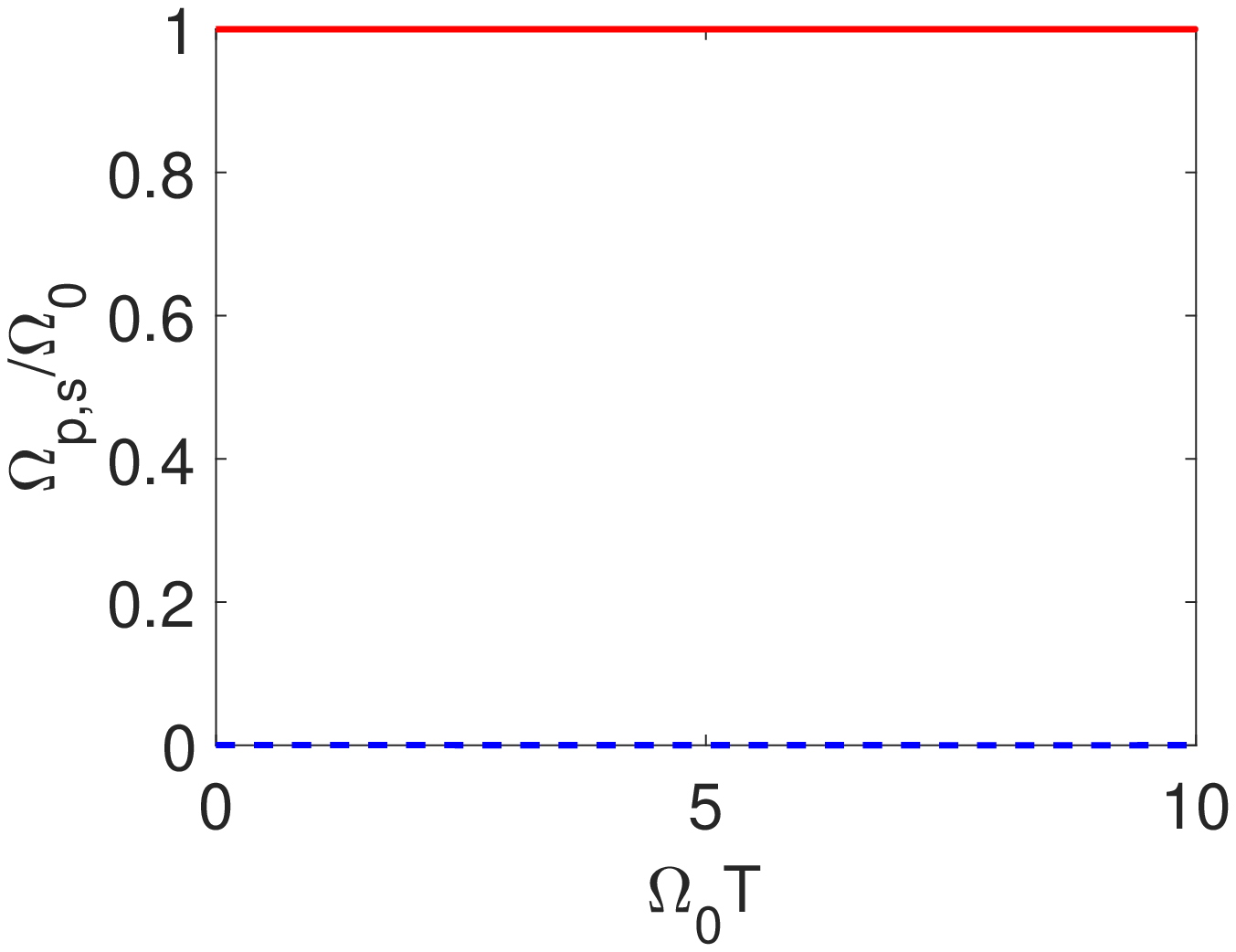}} &
       \subfigure[$\ $]{
	            \label{fig:pop_2_10}
	            \includegraphics[width=.45\linewidth]{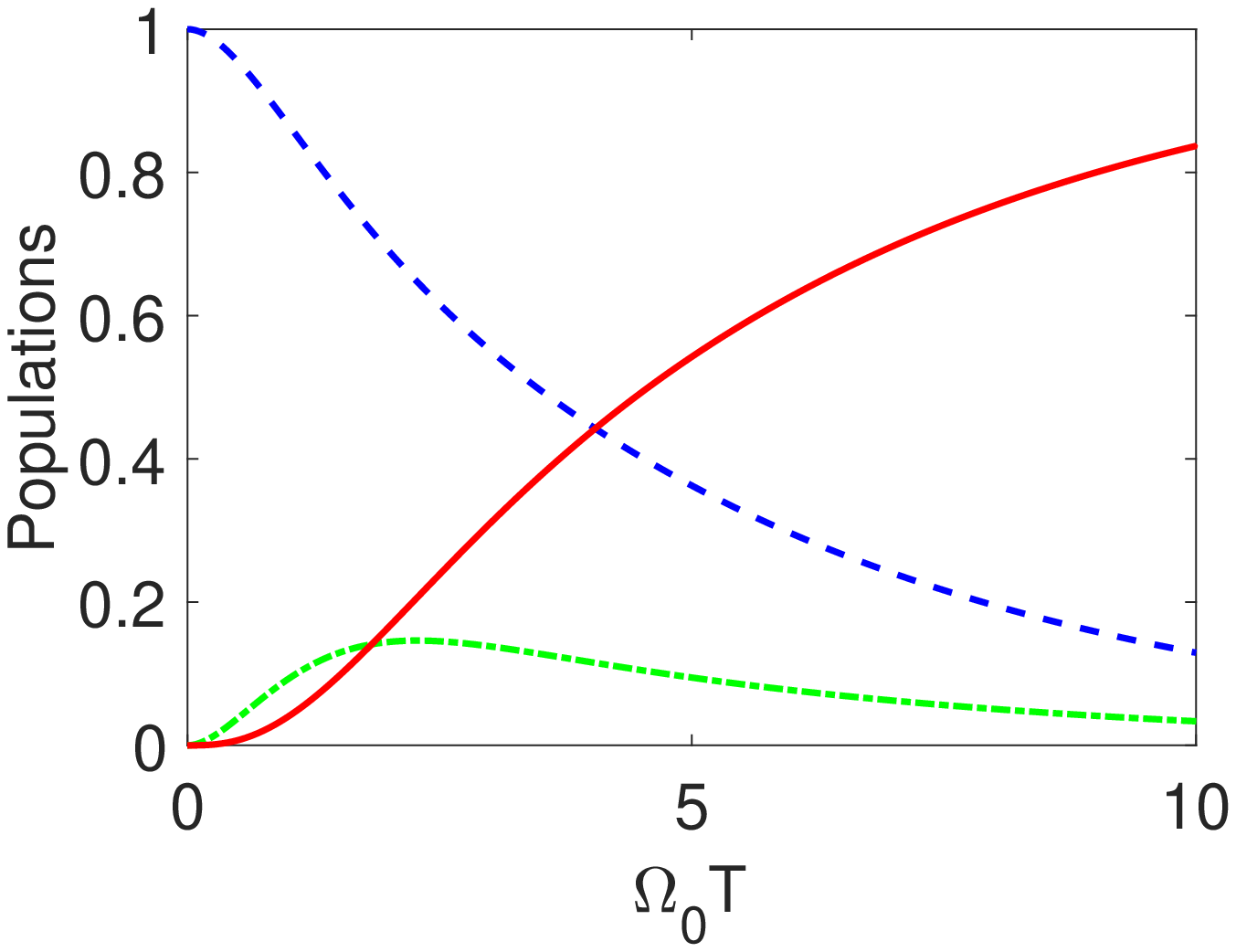}} \\
       \subfigure[$\ $]{
	            \label{fig:con_2_20}
	            \includegraphics[width=.45\linewidth]{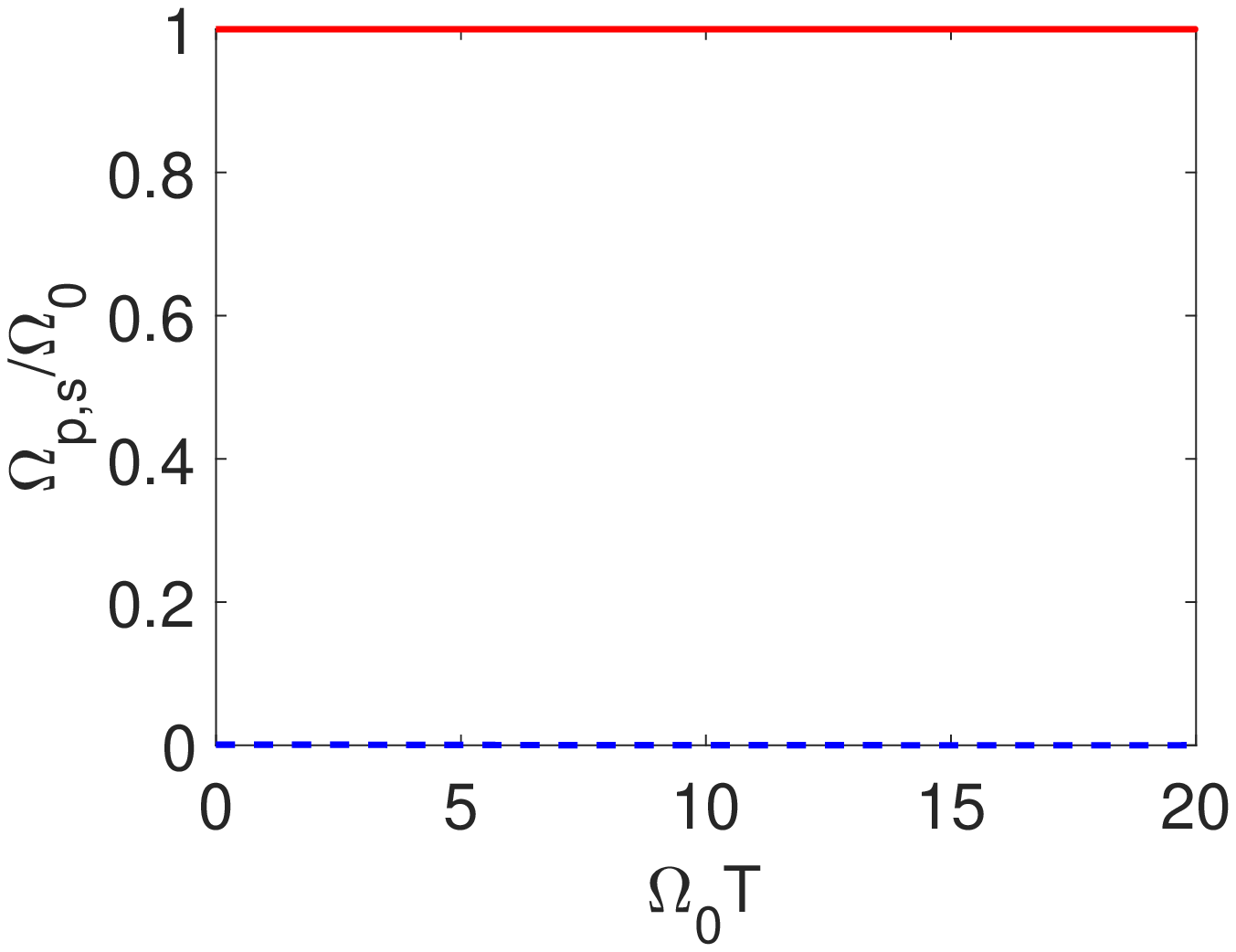}} &
       \subfigure[$\ $]{
	            \label{fig:pop_2_20}
	            \includegraphics[width=.45\linewidth]{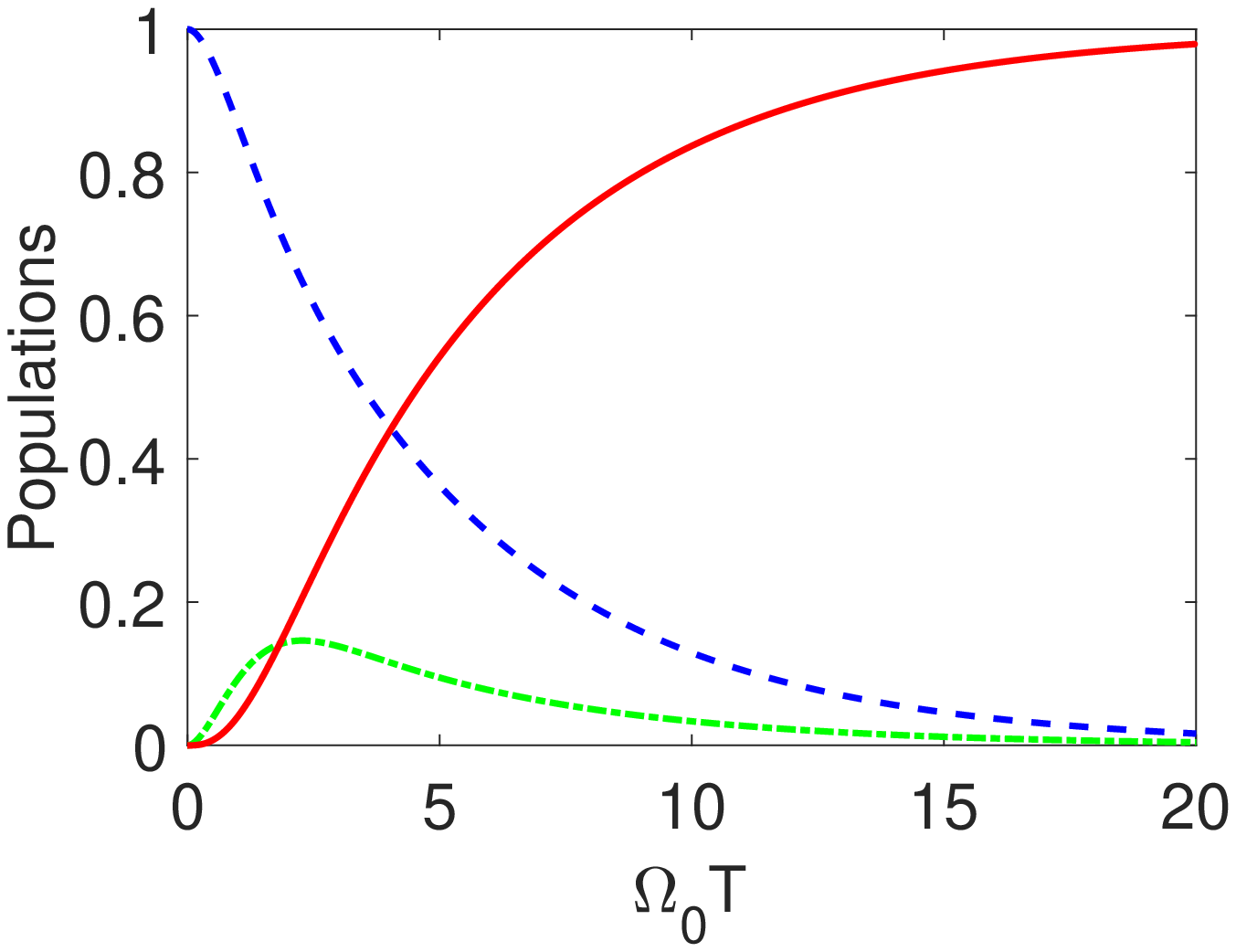}} \\
       \subfigure[$\ $]{
	            \label{fig:con_2_40}
	            \includegraphics[width=.45\linewidth]{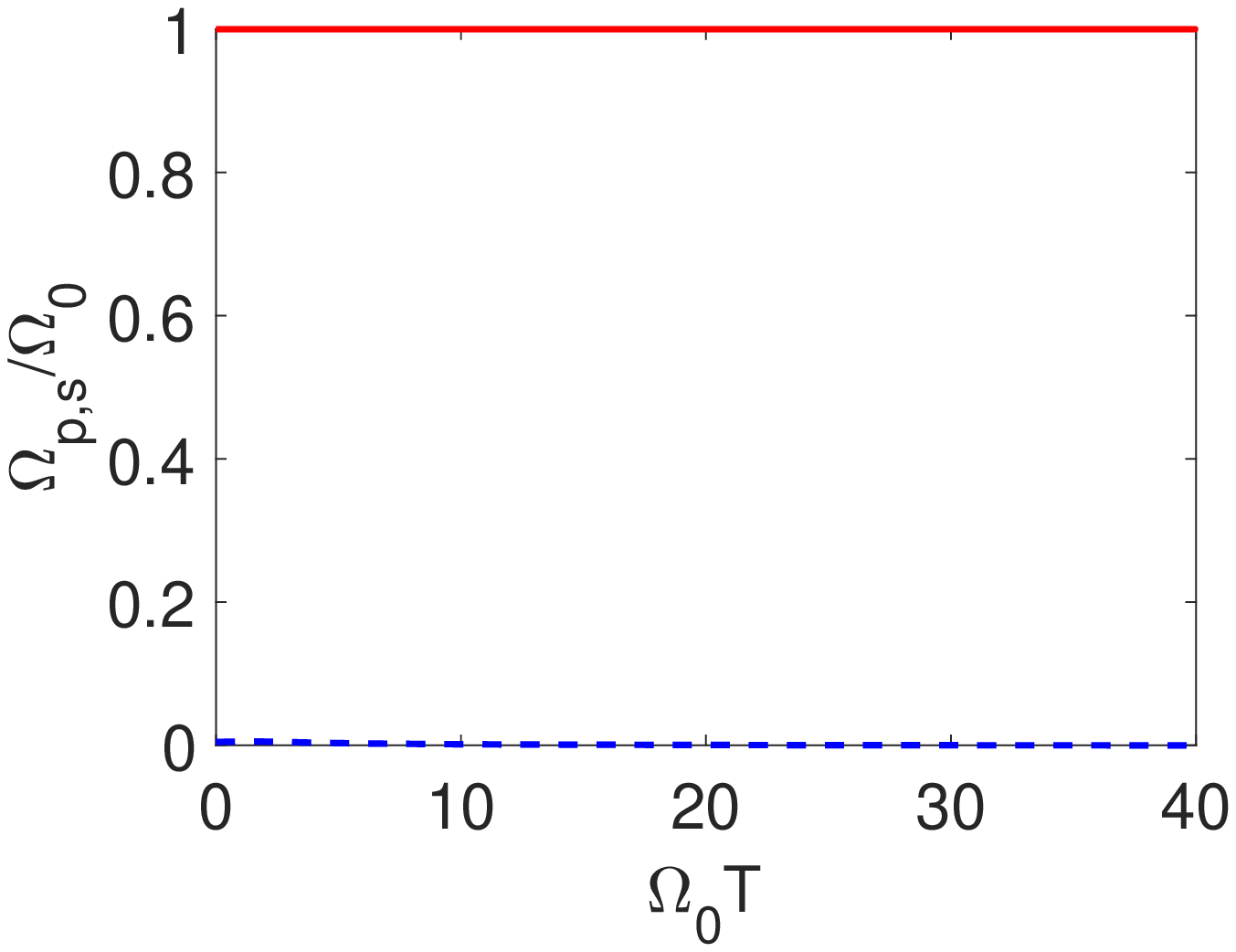}} &
       \subfigure[$\ $]{
	            \label{fig:pop_2_40}
	            \includegraphics[width=.45\linewidth]{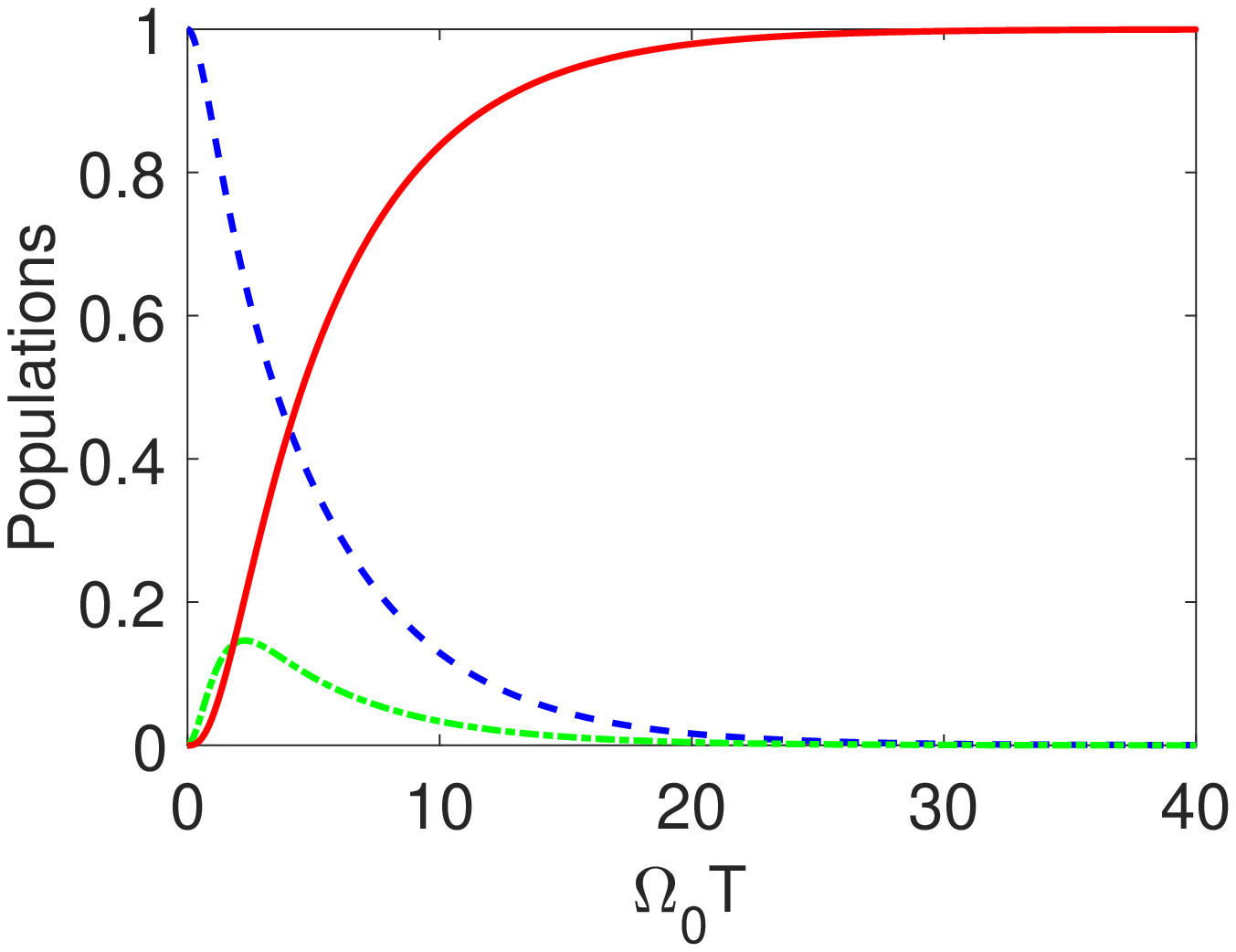}} \\
		\end{tabular}
\caption{(a, c, e) Optimal controls $\Omega_p(t)$ (red solid curve) and $\Omega_s(t)$ (blue dashed curve) obtained with numerical optimization for decay rate $\Gamma/\Omega_0=2$ and three different durations $\Omega_0T=10, 20, 40$ from top to bottom. (b, d, f) Corresponding evolution of populations $\rho_{11}(t)$ (blue dashed curve), $\rho_{22}(t)$ (green dashed-dotted curve), and $\rho_{33}(t)$ (red solid curve).}
\label{fig:example2}
\end{figure*}

\begin{figure*}[t]
 \centering
		\begin{tabular}{cc}
     	\subfigure[$\ $]{
	            \label{fig:con_2_10}
	            \includegraphics[width=.45\linewidth]{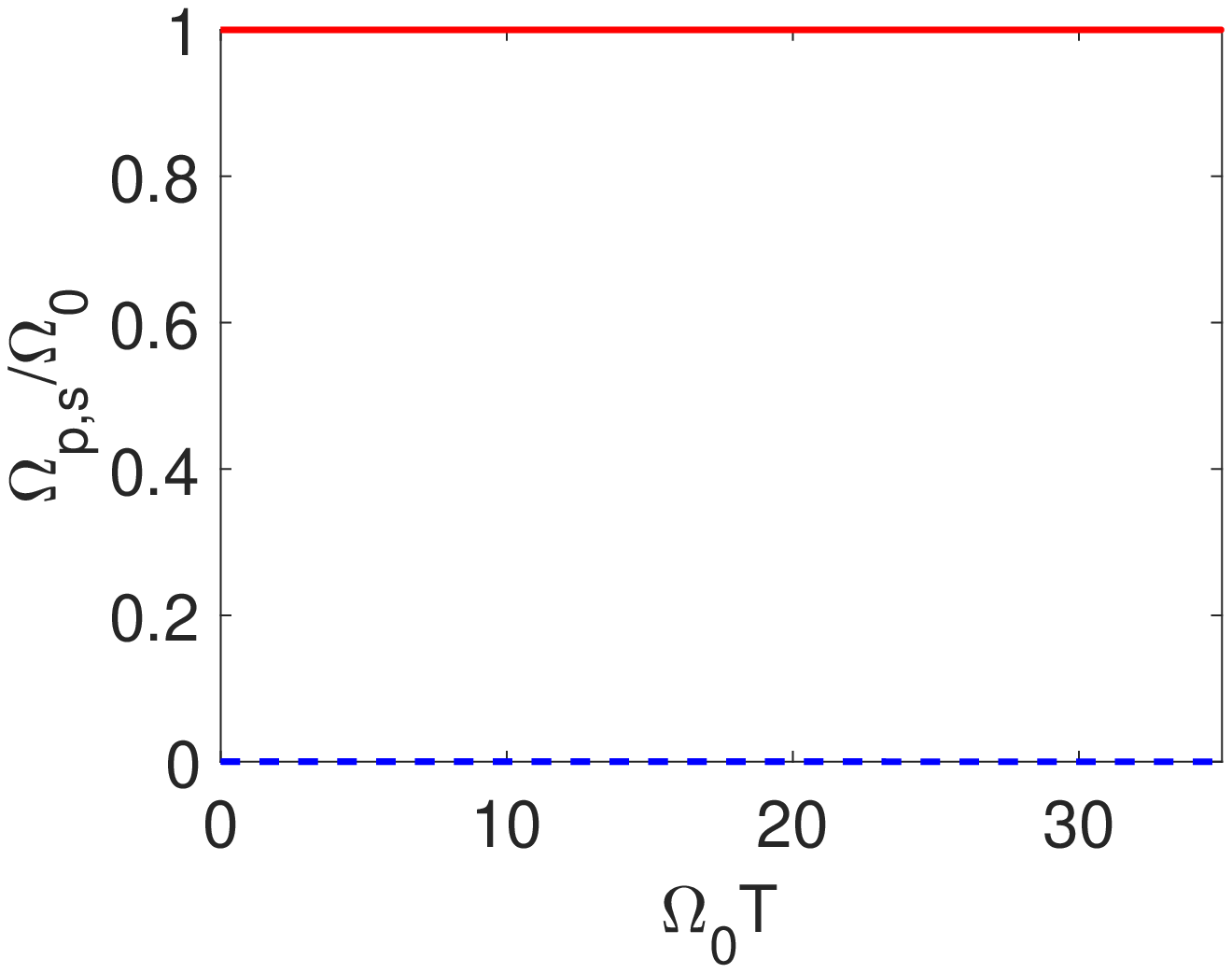}} &
       \subfigure[$\ $]{
	            \label{fig:pop_2_10}
	            \includegraphics[width=.45\linewidth]{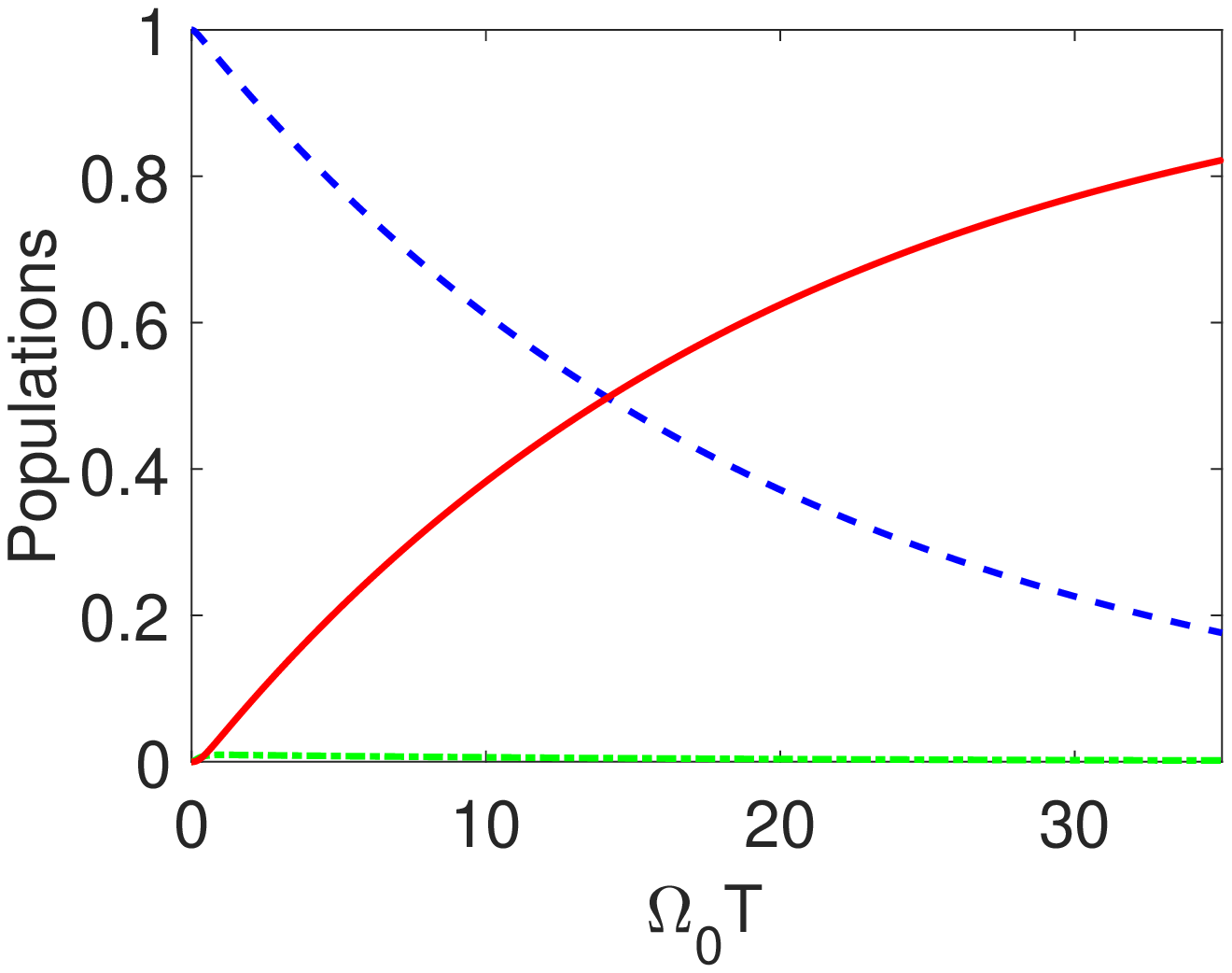}} \\
       \subfigure[$\ $]{
	            \label{fig:con_2_20}
	            \includegraphics[width=.45\linewidth]{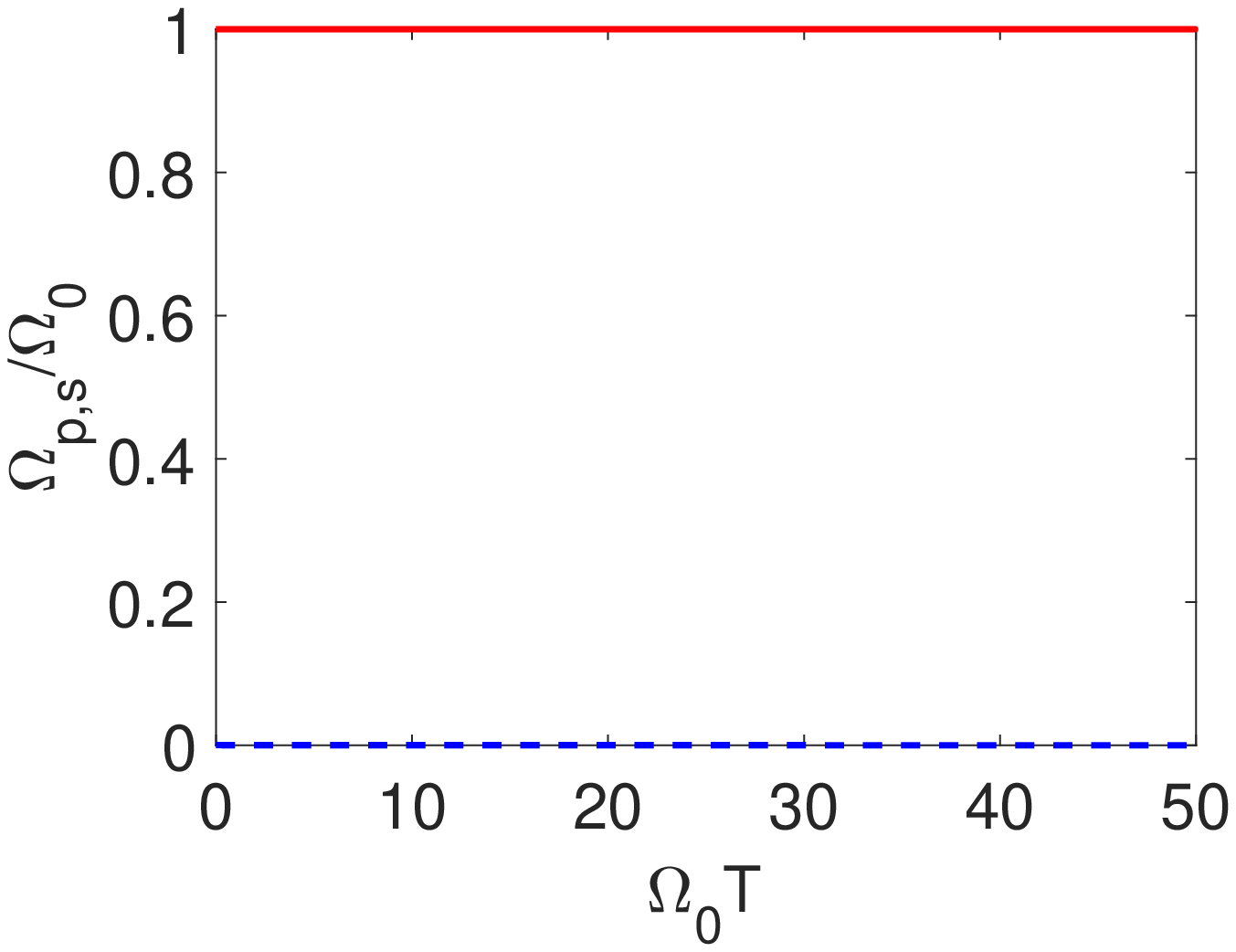}} &
       \subfigure[$\ $]{
	            \label{fig:pop_2_20}
	            \includegraphics[width=.45\linewidth]{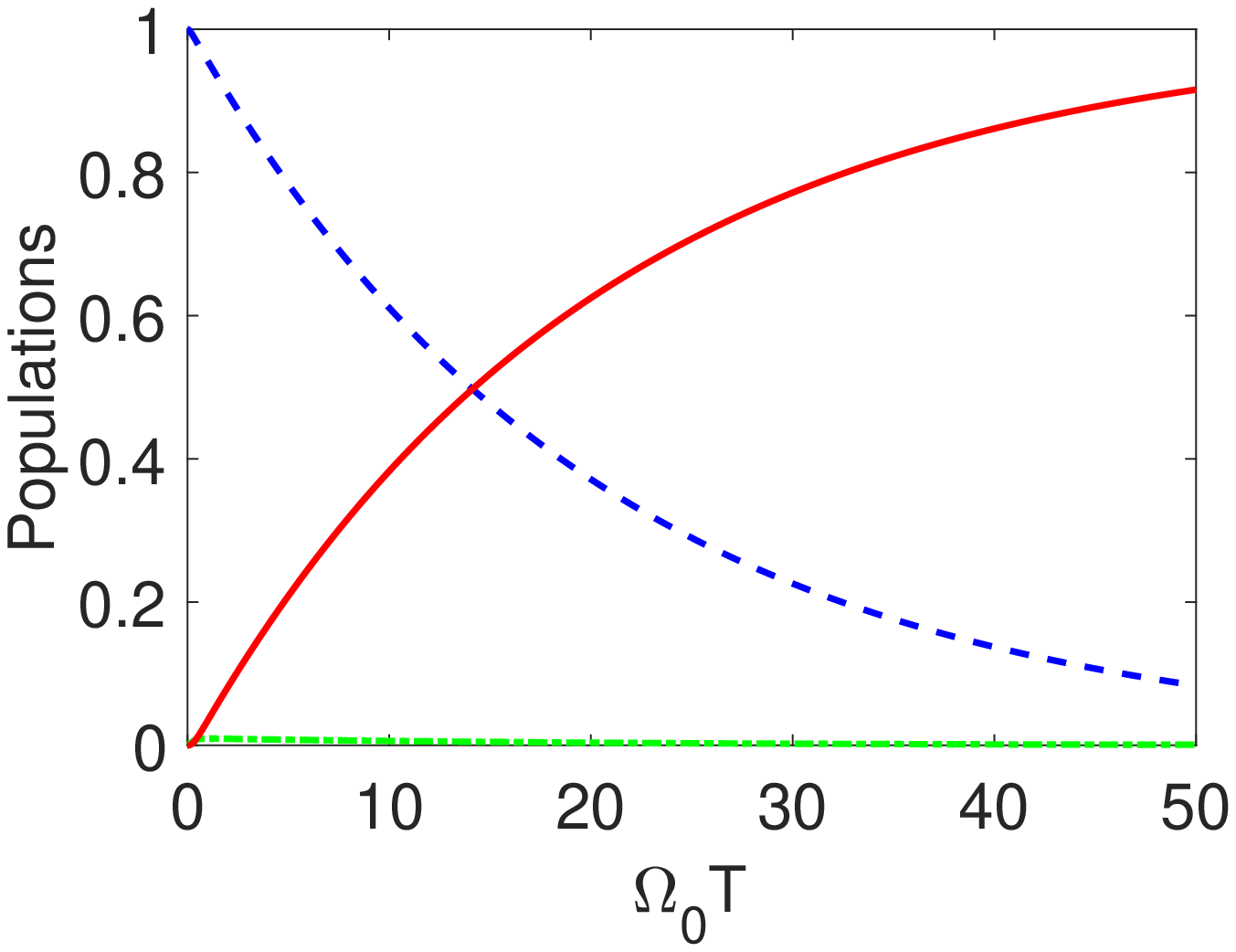}} \\
       \subfigure[$\ $]{
	            \label{fig:con_2_40}
	            \includegraphics[width=.45\linewidth]{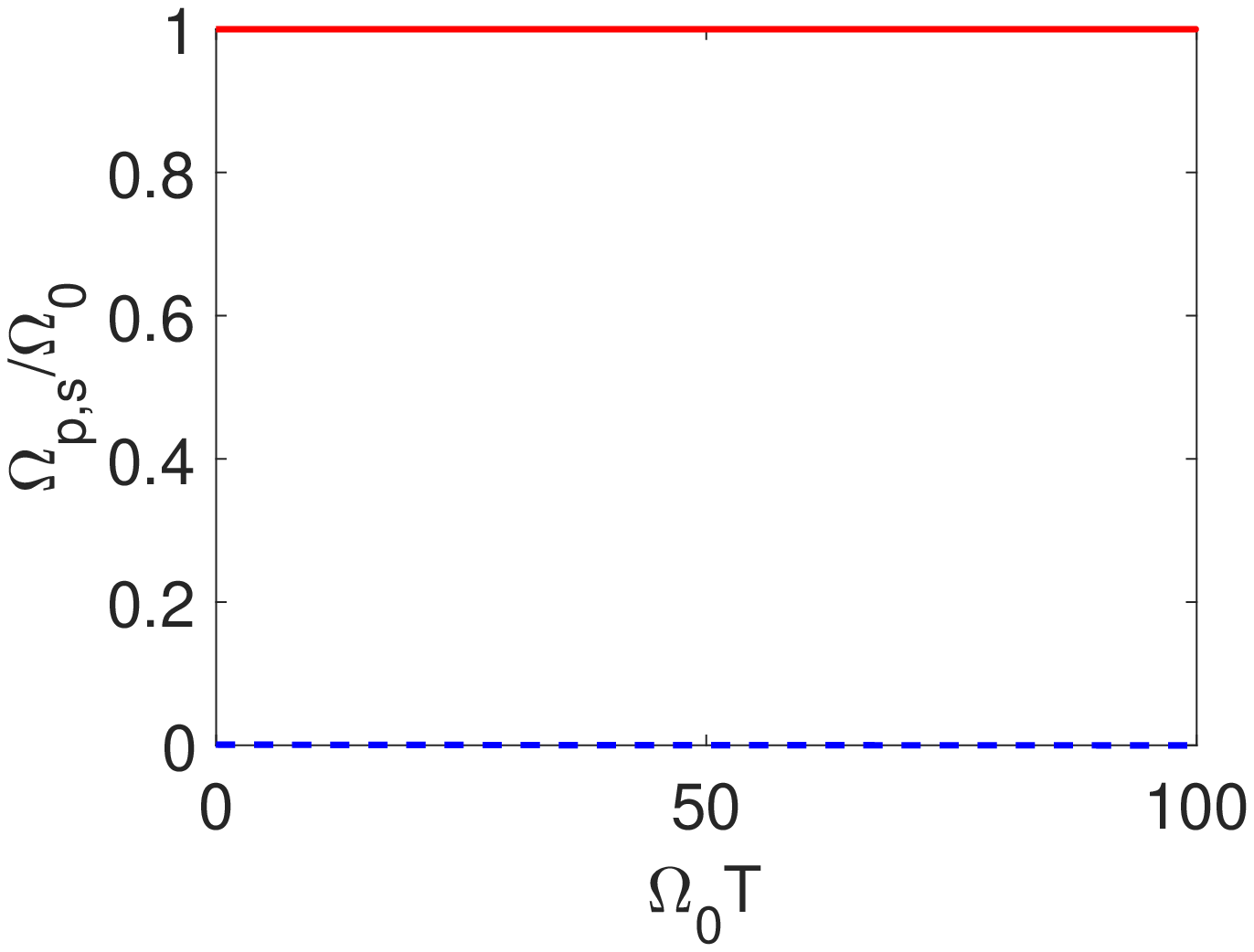}} &
       \subfigure[$\ $]{
	            \label{fig:pop_2_40}
	            \includegraphics[width=.45\linewidth]{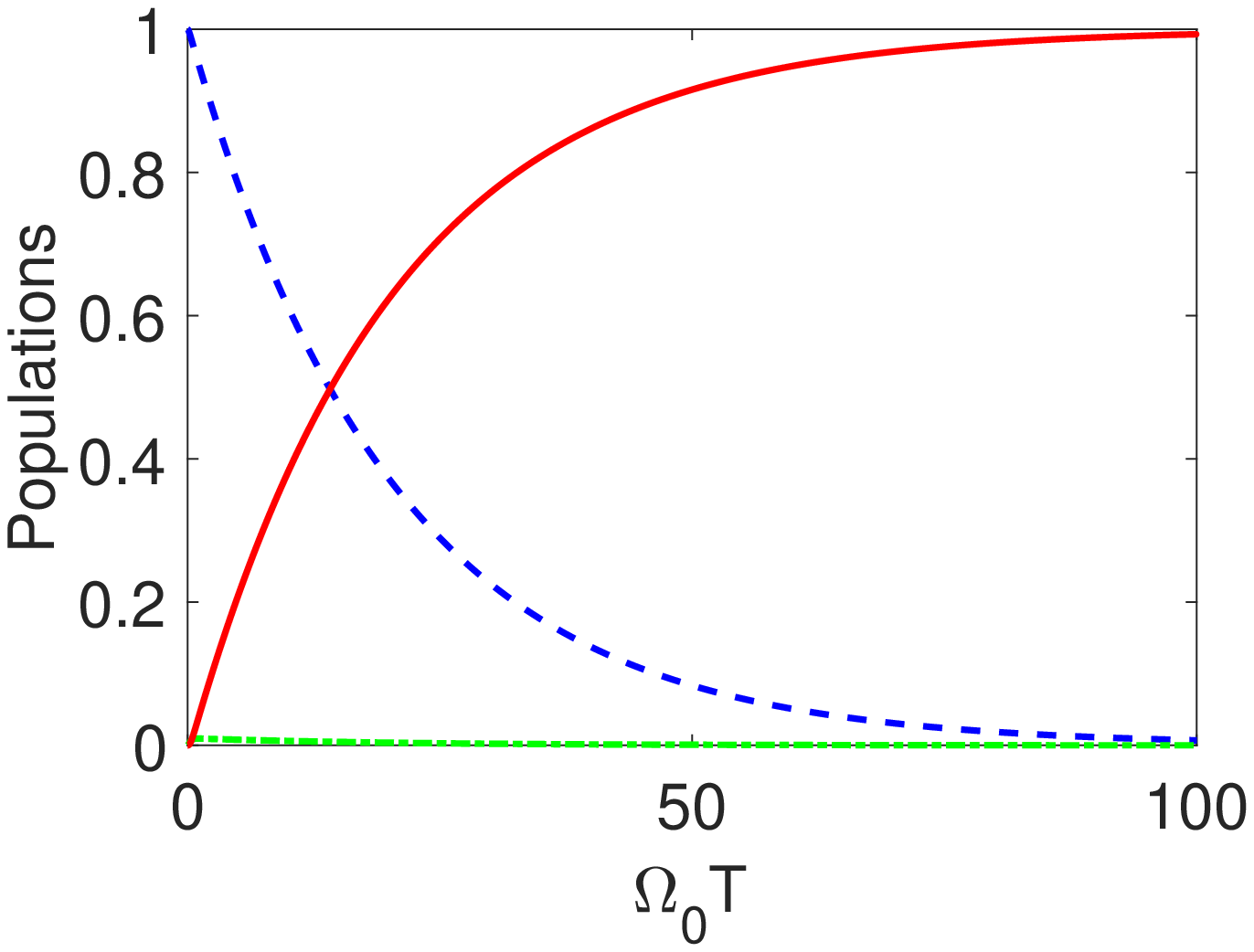}} \\
		\end{tabular}
\caption{(a, c, e) Optimal controls $\Omega_p(t)$ (red solid curve) and $\Omega_s(t)$ (blue dashed curve) obtained with numerical optimization for decay rate $\Gamma/\Omega_0=10$ and three different durations $\Omega_0T=35, 50, 100$ from top to bottom. (b, d, f) Corresponding evolution of populations $\rho_{11}(t)$ (blue dashed curve), $\rho_{22}(t)$ (green dashed-dotted curve), and $\rho_{33}(t)$ (red solid curve).}
\label{fig:example3}
\end{figure*}

In order to find the optimal controls $\Omega_p(t), \Omega_s(t)$ which maximize $x_3(T)=\rho_{33}(T)$ while satisfy constraint (\ref{constraint}) we use the optimal control solver BOCOP \cite{bocop}, which can easily incorporate such constraints. Note that the same results are obtained if we use the inequality constraint $\Omega^2_p(t)+\Omega^2_s(t)\leq\Omega_0^2$ instead of (\ref{constraint}), since the optimal pulses need to be ``filled" in order to fully exploit the available duration. In the first column of Fig. \ref{fig:example1} we display the numerically obtained optimal controls $\Omega_p$ (red solid curve) and $\Omega_s$ (blue dashed curve) for $\Gamma/\Omega_0=0.1$ and three different durations $\Omega_0T=5, 10, 20$, from top to bottom. In the second column we display the corresponding evolution of populations $\rho_{11}(t)$ (blue dashed curve), $\rho_{22}(t)$ (green dashed-dotted curve), and $\rho_{33}(t)$ (red solid curve). Observe that for this small decay rate the optimal controls deviate from optical pumping, which in the previous section we proved to be optimal but for large decays. In Fig. \ref{fig:example2} we display similar results for the moderate decay rate $\Gamma/\Omega_0=2$ and durations $\Omega_0T=10, 20, 40$, from top to bottom. Now observe that the numerically derived pulses conform to the optical pumping scheme, with almost all the available energy concentrated in the pump pulse, while the Stokes pulse is close to zero. Similar results are also obtained for the larger decay rate $\Gamma/\Omega_0=10$, shown in Fig. \ref{fig:example3}. These results confirm the optimality of optical pumping in the case of large decay rates.

\begin{figure*}[t]
 \centering
		\begin{tabular}{cc}
     	\subfigure[$\ $]{
	            \label{fig:con_a_n8}
	            \includegraphics[width=.40\linewidth]{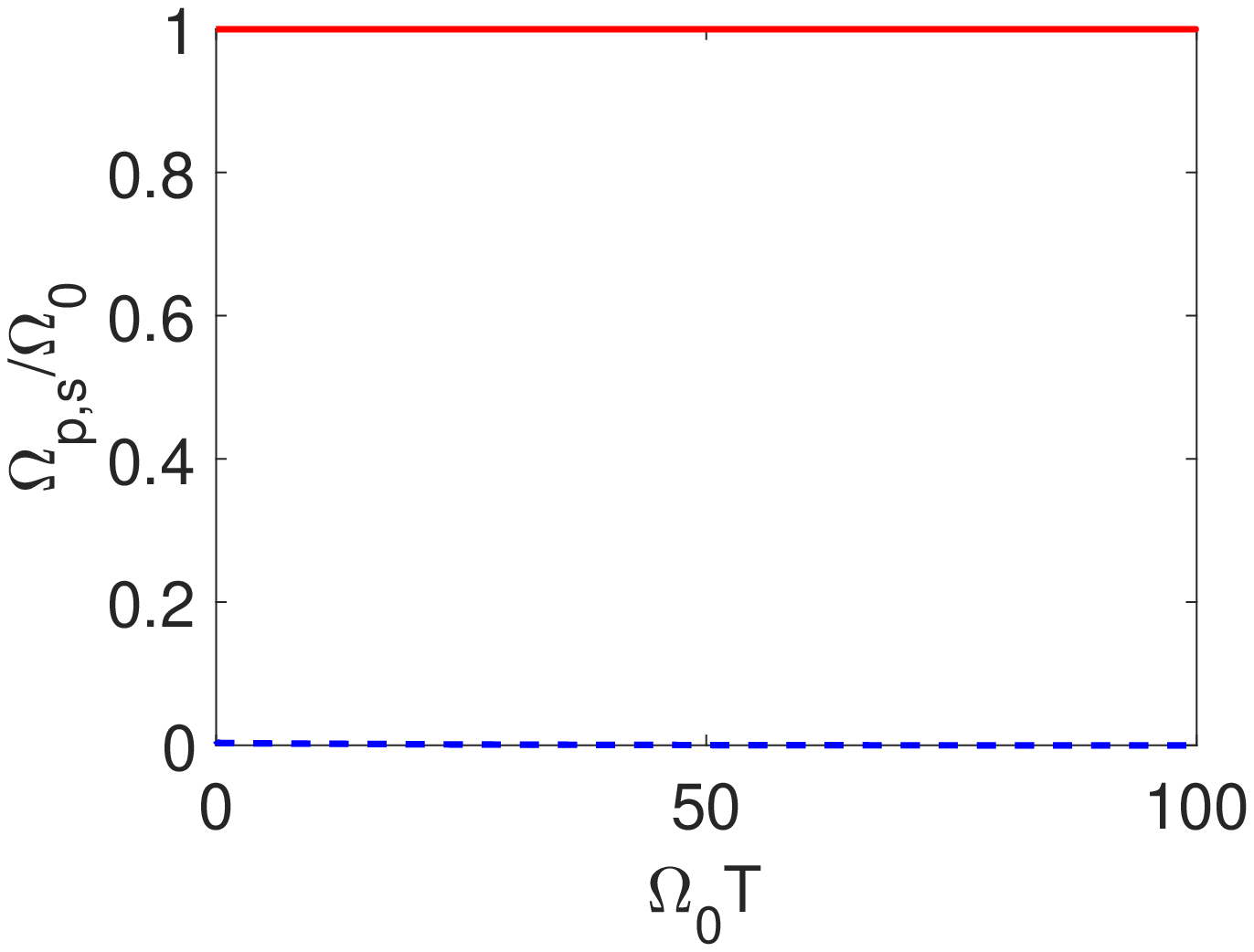}} &
       \subfigure[$\ $]{
	            \label{fig:pop_a_n8}
	            \includegraphics[width=.40\linewidth]{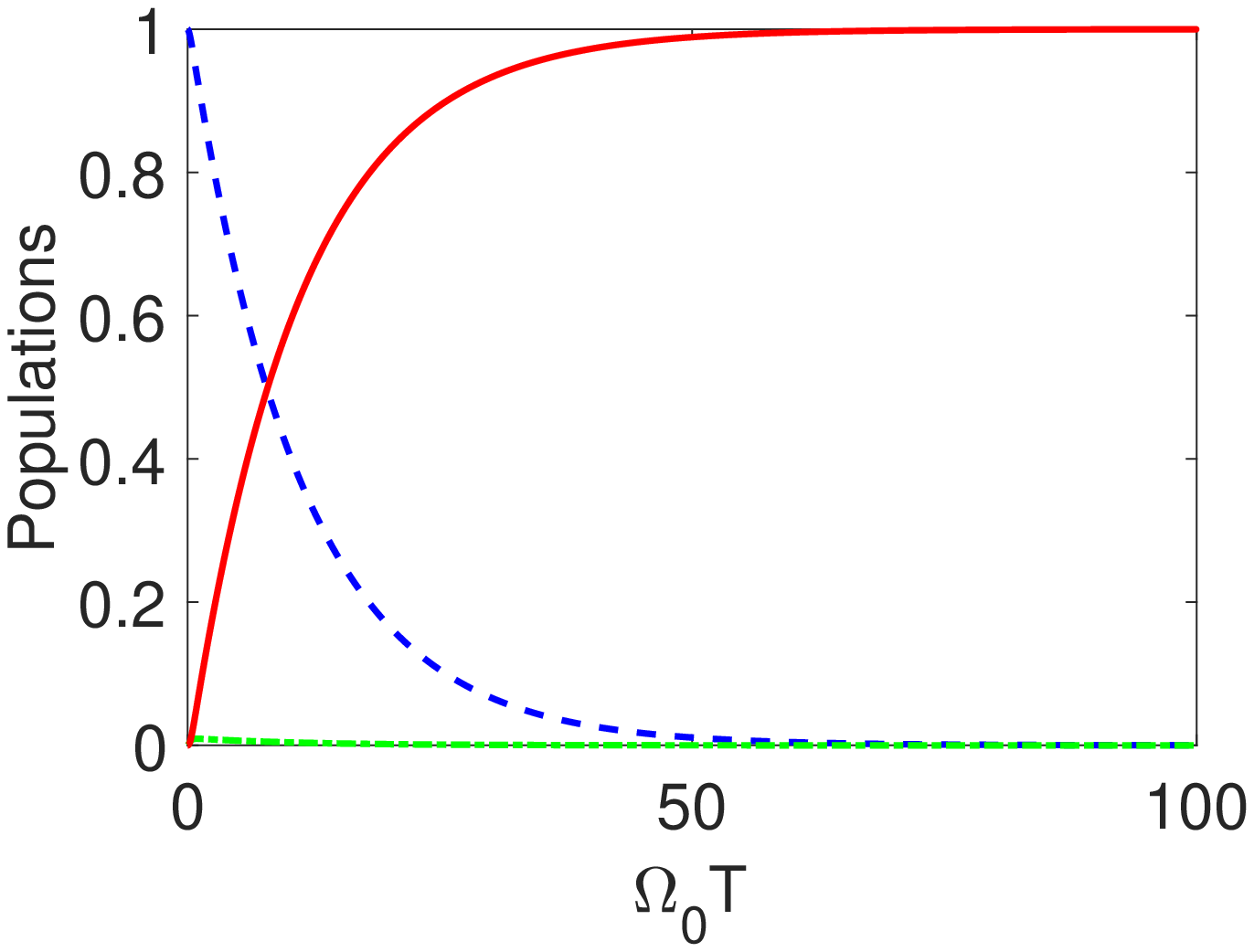}} \\
       \subfigure[$\ $]{
	            \label{fig:con_a_n2}
	            \includegraphics[width=.40\linewidth]{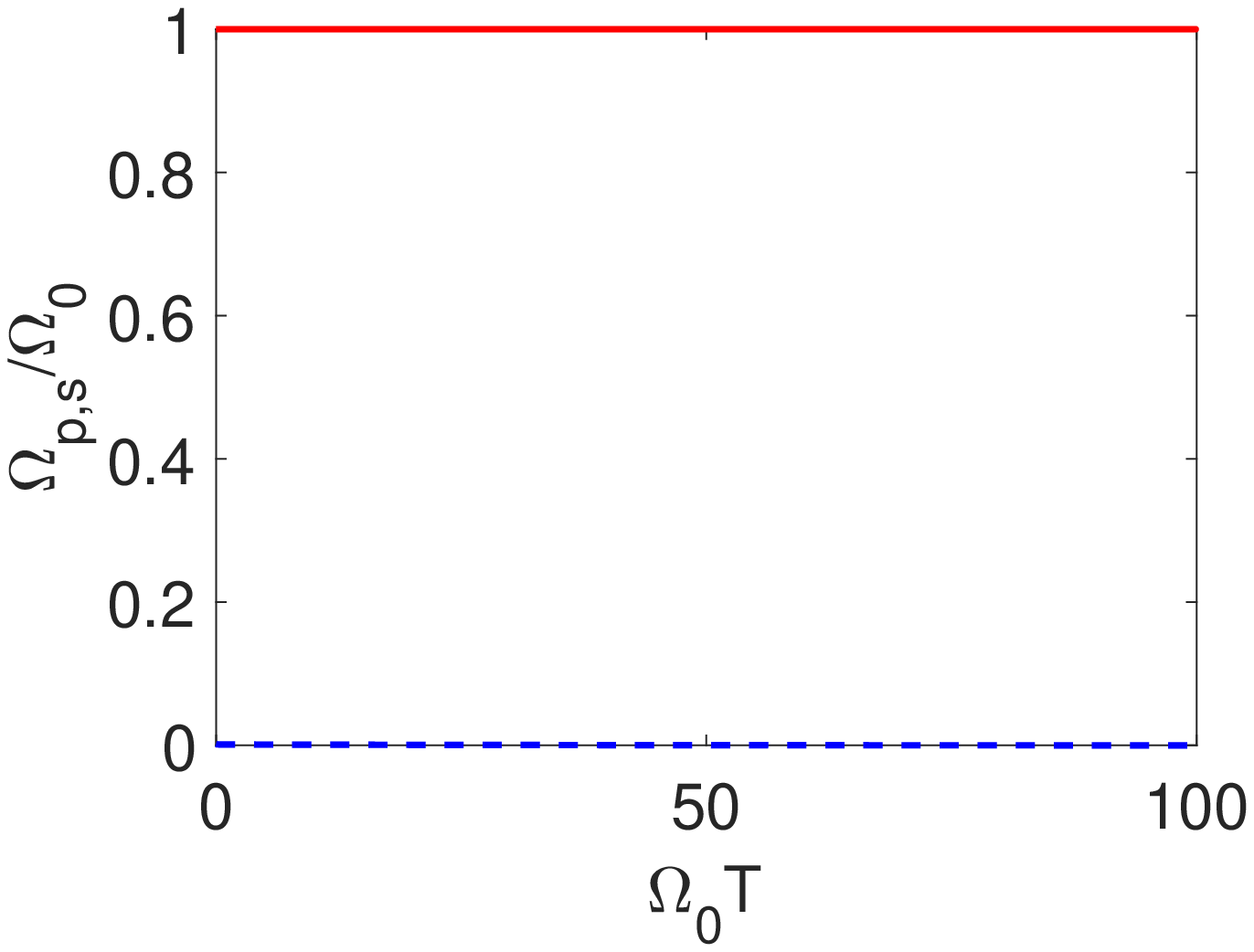}} &
       \subfigure[$\ $]{
	            \label{fig:pop_a_n2}
	            \includegraphics[width=.40\linewidth]{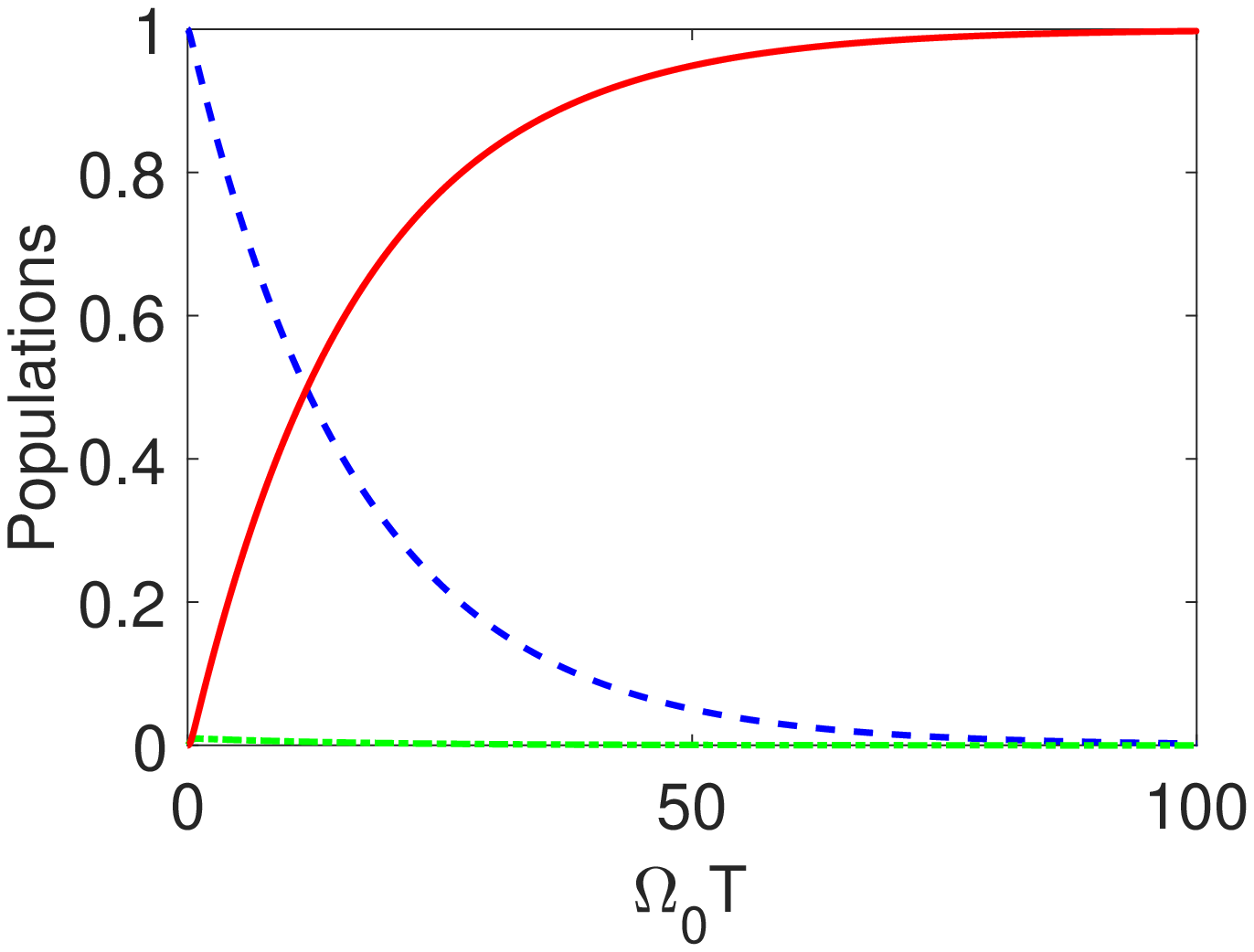}} \\
       \subfigure[$\ $]{
	            \label{fig:con_a_p2}
	            \includegraphics[width=.40\linewidth]{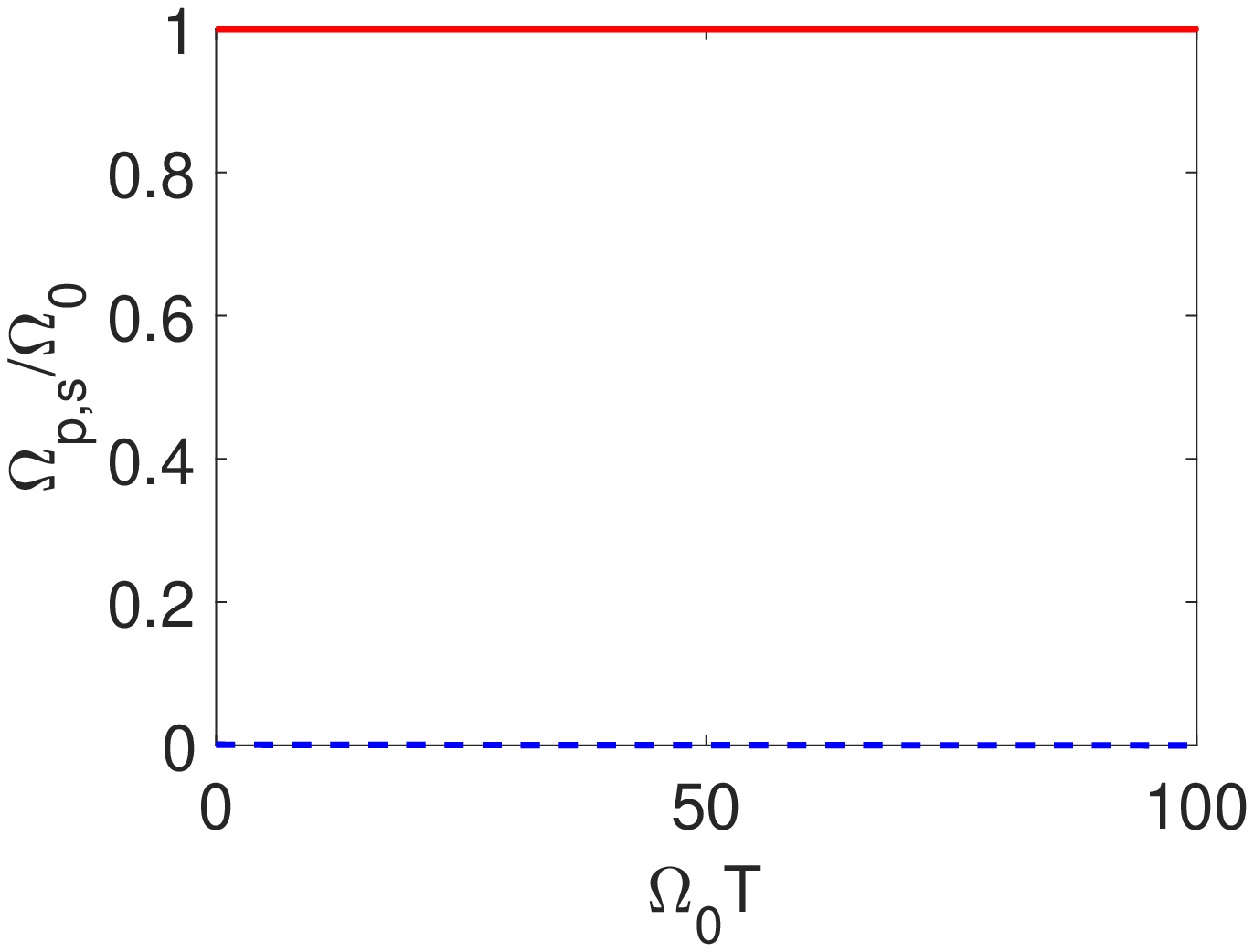}} &
       \subfigure[$\ $]{
	            \label{fig:pop_a_p2}
	            \includegraphics[width=.40\linewidth]{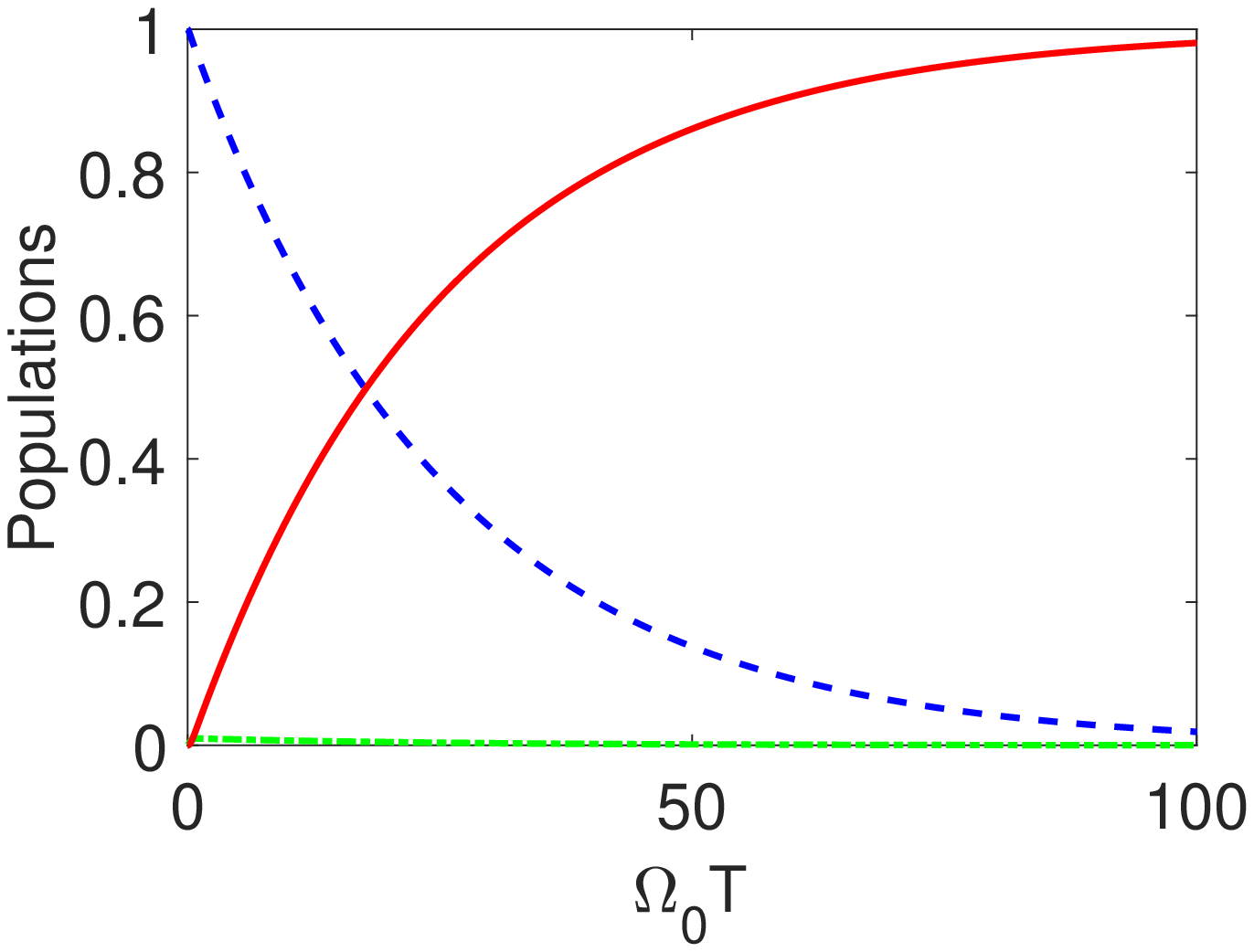}} \\
       \subfigure[$\ $]{
	            \label{fig:con_a_p8}
	            \includegraphics[width=.40\linewidth]{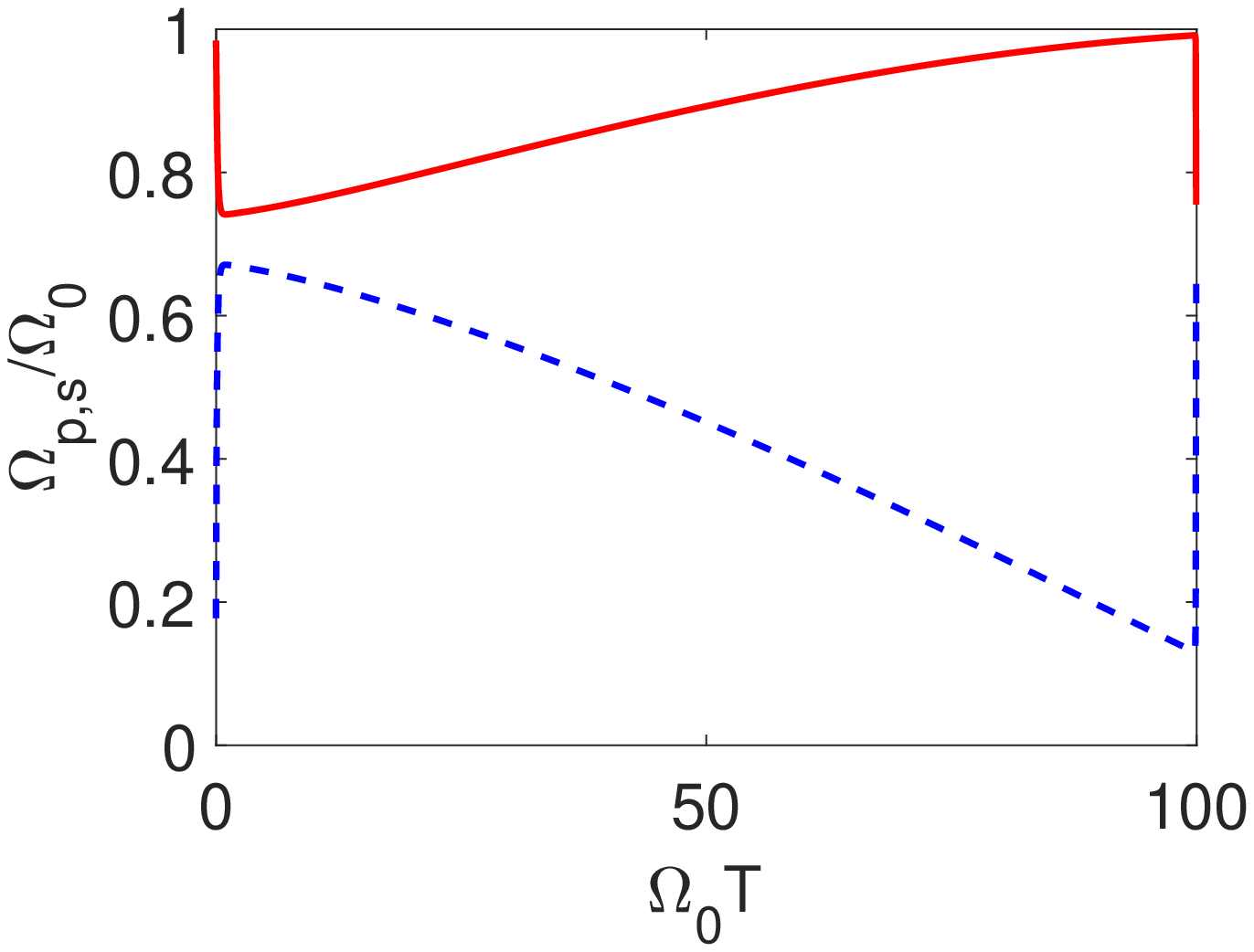}} &
       \subfigure[$\ $]{
	            \label{fig:pop_a_p8}
	            \includegraphics[width=.40\linewidth]{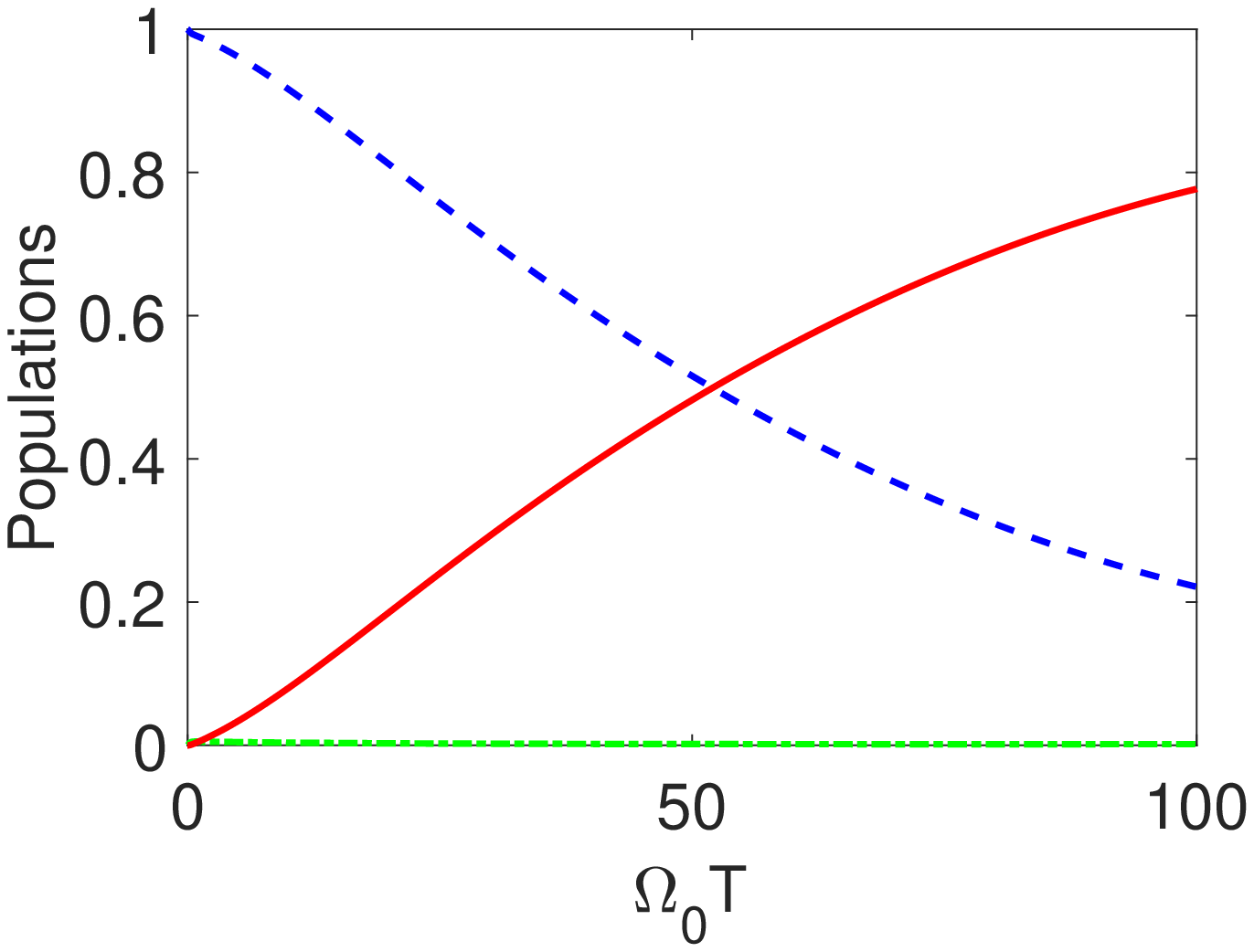}} \\
		\end{tabular}
\caption{Numerically obtained optimal controls $\Omega_p(t)$ (red solid curve), $\Omega_s(t)$ (blue dashed curve) and corresponding evolution of populations $\rho_{11}(t)$ (blue dashed curve), $\rho_{22}(t)$ (green dashed-dotted curve) and $\rho_{33}(t)$ (red solid curve), for fixed $\Gamma/\Omega_0=(\Gamma_1+\Gamma_3)/\Omega_0=10$, $\Omega_0T=100$, and four different values of $\gamma=\Gamma_1-\Gamma_3$: (a, b) $\gamma/\Omega_0=-8$, (c, d) $\gamma/\Omega_0=-2$, (e, f) $\gamma/\Omega_0=2$, (g, h) $\gamma/\Omega_0=8$.}
\label{fig:example4}
\end{figure*}

Up to now we have considered the symmetric case with equal dissipation rates from state $|2\rangle$ to states $|1\rangle$ and $|3\rangle$, $\Gamma_1=\Gamma_3=\Gamma/2$. For $\Gamma_1\neq\Gamma_3$, the angle $\theta$ ceases to be a cyclic variable in Eqs. (\ref{y}), (\ref{x}) \cite{Ivanov05}, and as a consequence the analytical solution of the corresponding optimal control problem becomes very difficult, if not impossible. Nevertheless, we can still investigate the nonsymmetric case using numerical optimal control. Intuitively, we expect that for $\Gamma_1<\Gamma_3$, i.e when the decay rates favor the population transfer from $|1\rangle$ to $|3\rangle$, the optimal pumping scheme remains optimal. For $\Gamma_1>\Gamma_3$ we expect the optimality of optical pumping to persist when the decay rates are not very different from each other. If we set $\Gamma=\Gamma_1+\Gamma_3$ and $\gamma=\Gamma_1-\Gamma_3$ \cite{Ivanov05}, then $\Gamma_1=(\Gamma+\gamma)/2$ and $\Gamma_3=(\Gamma-\gamma)/2$. In order to take the asymmetry into account, in Eqs. (\ref{x1}) and (\ref{x3}) we just need to replace $\Gamma$ with $\Gamma+\gamma$ and $\Gamma-\gamma$, respectively. In Fig. \ref{fig:example4} we fix $(\Gamma_1+\Gamma_3)/\Omega_0=\Gamma/\Omega_0=10$ and $\Omega_0T=100$, while we use four different values of $\gamma/\Omega_0=(\Gamma_1-\Gamma_3)/\Omega_0=-8, -2, 2, 8$, one for each row. As in the previous figures, the first column displays the numerically obtained optimal pump and Stokes pulses, while the second column the corresponding evolution of populations. Observe that for $\gamma<0$ ($\Gamma_1<\Gamma_3$), optical pumping remains optimal for both large (Fig. \ref{fig:con_a_n8}) and small (Fig. \ref{fig:con_a_n2}) asymmetry. For $\gamma>0$ ($\Gamma_1>\Gamma_3$), the optimality of optical pumping is maintained for small asymmetry (Fig. \ref{fig:con_a_p2}), while it breaks down for large (Fig. \ref{fig:con_a_p8}). From the second column of Fig. \ref{fig:example4} and specifically the final population $\rho_{33}(T)$, observe also that the asymmetry $\Gamma_1<\Gamma_3$ favors the transfer from $|1\rangle$ and $|3\rangle$ \cite{Ivanov05}.

\section{Conclusion}

\label{sec:con}

We used optimal control theory and numerical optimal control to show that for a closed $\Lambda$-system where the excited state decays to the lower states with large equal rates, the optimal strategy for population transfer between the lower levels is optical pumping. We also showed numerically that optical pumping remains optimal when the decay rate to the target state is larger than that to the initial state or the two rates are not very different from each other. The current methodology can also be extended to open systems, as we did in our recent work \cite{Stefanatos21} for an open $\Lambda$-system with large decay in the intermediate level. A STIRAP-like optimal pulse-sequence was derived, with both pump and Stokes fields active. The current work may find application in various tasks of quantum information processing involving the studied systems.

\end{document}